\documentclass[apjl]{emulateapj}
\usepackage{apjfonts}
\usepackage{amsmath}
\usepackage{rotating}
\usepackage{lscape}

\def\arcsec{\hbox{$^{\prime\prime}$}}
\def\arcdeg{\mbox{$^\circ$}}%

\newcommand{\aapg}{\gtrsim}
\newcommand{\aapl}{\lesssim}
\newcommand{\cmjj}{\mbox{${\rm cm^{-2}}$}}
\newcommand{\etal}{et al.}
\newcommand{\hI}{\mbox{${\rm H\ I}$}}
\newcommand{\kms}{\mbox{km\ s${^{-1}}$}}
\newcommand{\lya}{\mbox{${\rm Ly}\alpha$}}

\begin{document}

\lefthead{Chen \etal}
\righthead{}

\slugcomment{Accepted for Publication in the Astrophysical Journal}

\title{High-Redshift Starbursting Dwarf Galaxies Revealed by $\gamma$-ray Burst Afterglows$^{1,2}$}
\author{HSIAO-WEN CHEN\altaffilmark{3},
DANIEL A.\ PERLEY\altaffilmark{4},
LINDSEY K.\ POLLACK\altaffilmark{5},
JASON X.\ PROCHASKA\altaffilmark{5}, 
JOSHUA S.\ BLOOM\altaffilmark{4,11}, 
MIROSLAVA DESSAUGES-ZAVADSKY\altaffilmark{6}, 
MAX PETTINI\altaffilmark{7},
SEBASTIAN LOPEZ\altaffilmark{8},
ALDO DALL'AGLIO\altaffilmark{9}, and
GEORGE D.\ BECKER\altaffilmark{10}
}

\altaffiltext{1}{Based in part on observations made with the NASA/ESA
Hubble Space Telescope, obtained at the Space Telescope Science
Institute, which is operated by the Association of Universities for
Research in Astronomy, Inc., under NASA contract NAS 5-26555. }

\altaffiltext{2}{Observations reported here were obtained in part at the
Magellan telescopes, a collaboration between the Observatories of the Carnegie
Institution of Washington, University of Arizona, Harvard University,
University of Michigan, and Massachusetts Institute of Technology.}

\altaffiltext{3}{Dept.\ of Astronomy \& Astrophysics and
Kavli Institute for Cosmological Physics,
University of Chicago, Chicago, IL, 60637, U.S.A. \\
{\tt hchen@oddjob.uchicago.edu}}

\altaffiltext{4}{Department of Astronomy, 601 Campbell Hall, University of
California, Berkeley, CA 94720 }

\altaffiltext{5}{UCO/Lick Observatory; University of California, Santa
  Cruz, Santa Cruz, CA 95064}

\altaffiltext{6}{Observatoire de Gen\`eve, 51 Ch. des Maillettes,
        1290 Sauverny, Switzerland}

\altaffiltext{7}{Institute of Astronomy, Madingley Rd., Cambridge, CB3 0HA, UK}

\altaffiltext{8}{Departamento de Astronom\'ia, Universidad de Chile, Casilla 36-D, Santiago, Chile}

\altaffiltext{9}{Astrophysikalisches Institut Potsdam, An der Sternwarte 16, D-14482 Potsdam, Germany}

\altaffiltext{10}{Carnegie Observatories, 813 Santa Barbara St., Pasadena, CA 91101}

\altaffiltext{11}{Sloan Research Fellow}


\begin{abstract}

We present a study of 15 long-duration $\gamma$-ray burst (GRB) host
galaxies at $z>2$.  The GRBs are selected with available early-time
afterglow spectra in order to compare interstellar medium (ISM)
absorption-line properties with stellar properties of the host
galaxies.  In addition to five previously studied hosts, we consider
new detections for the host galaxies of GRB\,050820 and GRB\,060206
and place 2-$\sigma$ upper limits to the luminosities of the remaining
unidentified hosts.  We examine the nature of the host galaxy
population and find that (1) the UV luminosity distribution of GRB
host galaxies is consistent with expectations from a UV luminosity
weighted random galaxy population with a median luminosity of $\langle
L(UV)\rangle=0.1\,L_*$; (2) there exists a moderate correlation
between UV luminosity and Si\,II $\lambda\,1526$ absorption width,
which together with the observed large line widths of $W(1526)> 1.5$
\AA\ for a large fraction of the objects suggests a galactic outflow
driven velocity field in the host galaxies; (3) there is tentative
evidence for a trend of declining ISM metallicity with decreasing
galaxy luminosity in the star-forming galaxy population at $z=2-4$;
(4) the interstellar UV radiation field is found $\approx
35-350\times$ higher in GRB hosts than the Galactic mean value; and
(5) additional galaxies are found at $\aapl 2''$ from the GRB host in
all fields with known presence of strong Mg\,II absorbers, but no
additional faint galaxies are found at $\aapl 2''$ in fields without
strong Mg\,II absorbers.  Our study confirms that the GRB host
galaxies (with known optical afterglows) are representative of
unobscured star-forming galaxies at $z>2$, and demonstrates that high
spatial resolution images are necessary for an accurate identification
of GRB host galaxies in the presence of strong intervening absorbers.

\end{abstract} 


\section{INTRODUCTION}

  Gamma-ray bursts (GRBs) are among the most energetic events in the
universe.  In particular, long-duration GRBs are believed to originate
in the catastrophic death of massive stars (e.g.\ Woosley 1993;
Paczy\'nski 1998; Bloom \etal\ 2002; Stanek \etal\ 2003; Hjorth \etal\
2003; see Woosley \& Bloom 2006 for a recent review).  Since massive
stars evolve rapidly, long-duration GRBs should probe instantaneous
star formation out to the highest redshifts (e.g.\ Wijers \etal\ 1998)
with the afterglows serving as signposts to starburst galaxies in the
distant universe.

  Many bursts are followed by optical afterglows that can briefly
exceed the absolute brightness of any known quasar by orders of
magnitude (e.g.\ Akerlof \etal\ 1999; Kann \etal\ 2007; Bloom \etal\
2008) and serve as bright background sources for probing intervening
gas along the line of sight.  Early-time, high-resolution spectroscopy
of GRB afterglows have revealed numerous absorption features produced
by ground-state and excited-state ions in the interstellar medium
(ISM) of the host galaxies (e.g.\ Vreeswijk \etal\ 2004; Prochaska,
Chen, \& Bloom 2006; Vreeswijk \etal\ 2007; D'Elia \etal\ 2008).
Detailed studies based on comparisons of absorption-line strengths
have yielded accurate constraints on the host ISM properties,
including gas density, temperature, chemical composition, and
kinematics of the GRB host environment (e.g.\ Fynbo \etal\ 2006;
Savaglio 2006; Prochaska \etal\ 2007a).  Specifically, roughly 50\% of
known GRBs at $z>2$ are found in ISM of neutral gas column density
$N(\hI)>10^{21}$ \cmjj\ (Jakobsson \etal\ 2006; Chen \etal\ 2007a). In
addition, the host ISM generally exhibit moderate chemical enrichment with a
median metallicity of $> 1/10$ solar, although with a substantial
scatter over the range from $1/100$ to $\sim 1/2$ solar values (Fynbo
\etal\ 2006a; Savaglio 2006; Prochaska \etal\ 2007a).  Comparisons of
different ionic abundances also show that there exists a large
differential depletion with $[{\rm Zn}/{\rm Fe}]> +0.6$ dex
and $\alpha/{\rm Fe} \aapg 0.4$ dex, confirming the presence of a large
amount of gas mass and a chemical enrichment history dominated by
massive stars (Savaglio 2006; Prochaska \etal\ 2007a).  Finally, there
is a lack of molecular gas despite the presence of a large $N(\hI)$
(Fynbo \etal\ 2006; Tumlinson \etal\ 2007).

  Interpretations of these absorption-line data are not
straightforward.  In particular, the observed low-metal content in the
GRB host ISM, as opposed to optically selected luminous starburst
galaxies (e.g.\ Shapley \etal\ 2004), can be explained if the
progenitor stars originate in the outskirts of a luminous galaxy or if
the host galaxies are underluminous and have on average lower
metallicities.  A local-galaxy analogue of this is seen with the
so-called luminosity--metallicity relation (e,g.\ Tremonti \etal\ 2004
for nearby galaxies).  Recently, Fynbo \etal\ (2008) considered both
scenarios and showed that the metallicity distribution of GRB hosts is
consistent with the expectation that these host galaxies represent a
weighted star-forming galaxy population according to the on-going star
formation rate (SFR).  In addition, the large atomic gas column
density in contrast to the lack of molecular gas may be due to either
an enhanced UV radiation field in the star-forming regions near the
GRB progenitor or a relatively low dust and metal content indicated in
the absorption-line data (e.g.\ Tumlinson \etal\ 2007; Whalen \etal\
2008).  Recent detections of the 2175-\AA\ dust absorption feature in
GRBs\,070802 (Kr\"uhler \etal\ 2008; Eliasdottir \etal\ 2008) and
080805 (Jakobsson \etal\ 2008) offer an important test for this
scenario.  Regardless, these issues have direct impact on our
understanding of both the GRB progenitors and star formation physics.
Supplemental imaging and spectroscopic observations of the host
galaxies are necessary for accurate interpretations of the
absorption-line measurements.

  The transient nature of optical afterglows allows deep imaging and
spectroscopic studies of galaxies close to the lines of sight,
including the hosts, when the afterglows disappear (c.f.\ M{\o}ller
\etal\ 2002a; Chen \& Lanzetta 2003 for searches of damped \lya\
absorbing galaxies along quasar sightlines).  At $z>2$, where accurate
ISM abundance measurements are available based on afterglow
absorption-line spectroscopy, only four host galaxies have been
unambiguously identified (see Savaglio \etal\ 2008 for a compilation).
Comparison studies between known ISM properties from afterglow
absorption spectroscopy and the observed morphology and luminosity of
the host galaxies have been published individually for GRB\,000926
(Castro \etal\ 2003), GRB\,011211 (Vreeswijk \etal\ 2006), GRB\,021004
(Fynbo \etal\ 2005), and GRB\,030323 (Vreeswijk \etal\ 2004).  The
four GRB host galaxies together show an order-of-magnitude scatter in
their rest-frame UV luminosity and ISM metallicity.

  At $z<2$, where the majority of common ISM absorption features occur
at ultraviolet wavelengths and spectroscopic observations become
challenging on the ground, $>40$ GRB host galaxies have been
identified in late-time imaging follow-up.  These known host galaxies
provide important insights for understanding the nature of GRB
progenitors at intermediate redshifts.  First, Chary \etal\ (2002)
compared the estimated SFR and total stellar mass for 12 GRB host
galaxies and found that the host galaxies have on average higher SFR
per unit stellar mass than local starburst galaxies.  Le Floc'h \etal\
(2003) examined the optical and near-infrared $R-K$ colors and
rest-frame $B$-band luminosity function of 15 GRB host galaxies at
$\langle z\rangle\approx 1$.
They found that these host galaxies have on average bluer colors and
fainter luminosity ($\langle L_{B}\rangle \approx 0.1\,L_*$) than
random star-forming galaxies at the same redshift range.  Additional
mid-infrared imaging observations by these authors showed that this is
not due to dust extinction (Le Floc'h \etal\ 2006).  Similarly,
Christensen \etal\ (2004) studied the optical and near-infrared
spectral energy distributions (SEDs) of ten GRB host galaxies at
$\langle z\rangle=0.85$.
They found based on a comparison with $\sim 1000$ galaxies identified
at a similar redshift range in the Hubble Deep Fields that GRB host
galaxies have on average younger stellar age and shorter
characteristic star-forming time scale.  In addition, Fruchter \etal\
(2006) compared HST images of $>40$ GRB host galaxies at $\langle
z\rangle\approx 1$ with the host galaxies of core-collapse supernovae
(SNe).  They found that the majority of GRB host galaxies exhibit
irregular morphology, unlike the host galaxies of core-collapse SNe.
Recently, independent studies by Castro Cer\'on \etal\ (2008) and
Savaglio \etal\ (2008) show that $\langle z\rangle\aapl 1$ GRB host
galaxies contain on average lower stellar mass than field star-forming
galaxies.  Together, these results show that GRB host galaxies at
$\langle z\rangle\approx 1$ represent a relatively young dwarf
population that have experienced recent on-going star-forming
episodes.

  While GRB host galaxies based on the intermediate-redshift sample
appear to be underluminous and low mass systems, it is not clear
whether the long-duration GRBs originate preferentially in relatively
metal dificient star-forming regions (c.f.\ Wolf \& Podsiadlowski
2007; Modjaz \etal\ 2008). A low-metallicity environment is favored by
popular progenitor models so that the progenitor star can preserve
high spin and a massive stellar core to produce a GRB (e.g.\ Hirschi
\etal\ 2005; Yoon \& Langer 2005; Woosley \& Heger 2006).
Chemical abundance measurements for $z\aapl 2$ galaxies have been based
primarily on emission line observations of associated H\,II regions.
A subset of the host galaxies at $z<1$ have been followed up
spectroscopically for measuring emission-line fluxes.  A mean
metallicity of roughly $1/4$ solar is found but with a large scatter
(e.g.\ Sollerman \etal\ 2005; Modjaz \etal\ 2008; Savaglio \etal\
2008).  The accuracy of emission-line-based abundance estimates
depends sensitively on the accuracy of the calibrations between
different line diagnostics (e.g.\ Kewly \& Dopita 2002; Skillman
\etal\ 2003; Kennicutt \etal\ 2003).  Whether or not there exists a
maximum metallicity for forming a GRB remains an open question.

  We have carried out an optical and near-infrared imaging survey of
fields around 15 GRBs at $z>2$.  The GRBs are selected to have
early-time afterglow spectra in order to compare ISM absorption-line
properties with stellar properties.  The goal is to identify the host
galaxies and constrain their rest-frame UV and optical luminosities.
The primary objectives are (1) to quantify the luminosity distribution
of the GRB host galaxy populations and investigate whether or not the
GRB host galaxies trace the typical star-forming galaxies at high
redshift and (2) to examine whether there exists a correlation between
the ISM metal content and host luminosity.  The starburst nature of
GRB hosts makes this galaxy sample a unique laboratory for studying
star formation physics and stellar feedback at high redshift.  We
adopt a $\Lambda$CDM cosmology, $\Omega_{\rm M}=0.3$ and
$\Omega_\Lambda = 0.7$, with a dimensionless Hubble constant $h =
H_0/(100 \ {\rm km} \ {\rm s}^{-1}\ {\rm Mpc}^{-1})$ throughout the
paper.

\setcounter{footnote}{0} 

\section{The GRB SAMPLE}

\begin{center}
\begin{small}
\begin{deluxetable*}{p{1in}cccccccrc}
\tablewidth{0pt}
\tablecaption{Summary of the Optical and Near-infrared Imaging Data}
\tablehead{\multicolumn{1}{c}{Field} & \colhead{RA(J2000)} & \colhead{Dec(J2000)} & \colhead{$z_{\rm GRB}$} & \colhead{$\log\,N(\hI)$} & \colhead{References\tablenotemark{a}} & \colhead{Instrument} & \colhead{Filter} & \colhead{EXPTIME (s)} & \colhead{FWHM (\arcsec)}}
\startdata
GRB\,011211  \dotfill & 11:15:17.98 & $-$21:56:56.2 & 2.140 & $20.4\pm 0.2$ & (1) & HST/STIS\tablenotemark{b} & Clear & 19544 & 0.1 \nl
                      &             &               &       &  &  & Magellan/PANIC & $H$ & 12960 & 0.4 \nl
GRB\,020124  \dotfill & 09:32:50.81 & $-$11:31:10.6 & 3.198 & $21.7\pm 0.2$ & (2) & HST/STIS\tablenotemark{c} & Clear & 24798 & 0.1 \nl
                      &             &               &       &  &  & Magellan/PANIC & $H$ & 13860 & 0.4 \nl
GRB\,030323  \dotfill & 11:06:09.38 & $-$21:46:13.3 & 3.372 & $21.90\pm 0.07$ & (3) & HST/ACS/WFC\tablenotemark{d} & F606W & 5928 & 0.1 \nl
                      &             &               &       &  &  & Magellan/PANIC & $H$ & 14400 & 0.4 \nl
GRB\,030429  \dotfill & 12:13:07.50 & $-$20:54:49.7 & 2.658 & $21.6\pm 0.2$ & (4) & Magellan/PANIC & $H$ & 7740  & 0.4 \nl
GRB\,050401  \dotfill & 16:31:28.81 & $+$02:11:14.2 & 2.899 & $22.6 \pm 0.3$ & (5) & Magellan/PANIC & $H$ &  6300 & 0.6 \nl
                      &             &               &       &  &  & Keck/LRIS      & $g$ & 2540 & 1.1 \nl
                      &             &               &       &  &  & Keck/LRIS      & $R_c$ & 2460 & 1.0 \nl
GRB\,050730  \dotfill & 14:08:17.13 & $-$03:46:16.7 & 3.968 & $22.15 \pm 0.10$ & (6) &  Magellan/MagIC & $i'$ & 2700 & 0.6 \nl
                      &             &               &       &  &  & Keck/LRIS      & $g$ & 3900 & 1.2 \nl
                      &             &               &       &  &  & Keck/LRIS      & $R_c$ & 3900 & 1.0 \nl
GRB\,050820A \dotfill & 22:29:38.11 & $+$19:33:37.1 & 2.615 & $21.0 \pm 0.1 $ & (7) & HST/ACS/WFC\tablenotemark{e} & F625W & 2238 & 0.1 \nl
                      &             &               &       &  &  & HST/ACS/WFC & F775W & 4404 & 0.1 \nl
                      &             &               &       &  &  & HST/ACS/WFC & F850LP & 14280 & 0.1 \nl
                      &             &               &       &  &  & Magellan/PANIC & $H$ &  8460 & 0.6 \nl
                      &             &               &       &  &  & Keck/LRIS      & $g$ & 2620 & 0.7 \nl
                      &             &               &       &  &  & Keck/LRIS      & $R_c$ & 2500 & 1.9 \nl
GRB\,050908  \dotfill & 01:21:50.75 & $-$12:57:17.2 & 3.343 & $17.55 \pm 0.10$ & (8) & Magellan/PANIC & $H$ & 12150 & 0.5 \nl
GRB\,050922C \dotfill & 21:09:33.08 & $-$08:45:30.2 & 2.199 & $21.5 \pm 0.1$ & (7) & Magellan/PANIC & $H$ & 12560 & 0.5 \nl
                      &             &               &       &  &  & Keck/LRIS      & $g$ & 3870 & 1.2 \nl
                      &             &               &       &  &  & Keck/LRIS      & $R_c$ & 3660 & 1.0 \nl
GRB\,060206  \dotfill & 13:31:43.42 & $+$35:03:03.6 & 4.048 & $20.85 \pm 0.10$ & (9) & HST/ACS/WFC\tablenotemark{f} & F814W & 9886 & 0.1 \nl 
GRB\,060607  \dotfill & 21:58:50.40 & $-$22:29:46.7 & 3.075 & $16.85 \pm 0.10$ & (10) & Magellan/PANIC & $H$ & 18540 & 0.5 \nl
                      &             &               &       &  &  & Keck/LRIS      & $g$ & 2550 & 1.0 \nl
                      &             &               &       &  &  & Keck/LRIS      & $R_c$ & 2430 & 1.5 \nl
GRB\,070721B \dotfill & 02:12:32.97 & $-$02:11:40.4 & 3.626 & $21.50\pm 0.20$ & (11) & Magellan/PANIC & $H$ & 10320 & 0.5 \nl
\hline
\multicolumn{10}{c}{Previously Published Fields} \nl
\hline
GRB\,000301C \dotfill & 16:20:18.60 & $+$29:26:36.0 & 2.040 & $21.2\pm 0.5$ & (12) & HST/STIS\tablenotemark{g} & Clear & 16422 & 0.1 \nl
GRB\,000926  \dotfill & 17:04:09.62 & $+$51:47:11.2 & 2.038 & $21.3\pm 0.3$ & (13) & \multicolumn{4}{l}{Castro \etal\ (2003)} \nl
GRB\,021004  \dotfill & 00:26:54.68 & $+$18:55:41.6 & 2.329 & $19.5\pm 0.5$ & (14) & \multicolumn{4}{l}{Fynbo \etal\ (2005)}
\enddata
\tablenotetext{a}{(1): Vreeswijk \etal\ (2006); (2) Hjorth \etal\ (2003); (3) Vreeswijk \etal\ (2004); (4) Jakobsson \etal\ (2004); (5) Watson \etal\ (2006); (6) Chen \etal\ (2005); (7) Prochaska \etal\ (2007); (8) Fynbo, private communication; (9) Fynbo \etal\ (2006); (10) Chen \etal\ (2007), see also \S\ 4.11; (11) Malesani \etal\ (2007); (12) Jensen \etal\ (2001); (13) Fynbo \etal\ (2002); (14) Fynbo \etal\ (2005), but see \S\ 4.13 for discussion.}
\tablenotetext{b}{The imaging data were retrieved from the HST data archive; PID $= 8867$.  The host galaxy was clearly detected in these images that have been analyzed and published in Jakobsson \etal\ (2003).}
\tablenotetext{c}{The imaging data were retrieved from the HST data archive; PID $= 9180$.  A detailed analysis of these space images is published in Berger \etal\ (2002).  No emission from the host galaxy was found.}
\tablenotetext{d}{The imaging data were retrieved from the HST data archive; PID $= 9405$.  The host galaxy was found in the first-epoch images (exptime $=$ 1920 s), which have been analyzed and published by Vreeswijk \etal\ (2004).}
\tablenotetext{e}{The imaging data were retrieved from the HST data archive; PID $= 10551$.}
\tablenotetext{f}{The imaging data were obtained through our own program before ACS failed; PID $= 10817$.}
\tablenotetext{g}{The imaging data were retrieved from the HST data archive; PID $= 8189$.  The host galaxy was marginally detected in these images that have been analyzed and published in Fruchter \etal\ (2006).}
\end{deluxetable*}
\end{small}
\end{center}

  We generated a sample of 15 GRBs at $z_{\rm GRB}>2$ for studying the
nature of high-redshift GRB host galaxies and for comparing the host
properties with those of field galaxies in deep surveys.  The GRBs
were selected to have early-time, moderate-to-high resolution
afterglow spectra available for measuring the underlying neutral
hydrogen column density.  In addition, eight of the GRBs have
sufficiently high-spectral resolution for constraining the chemical
abundances of the host ISM.  Only five of these fields have late-time
images published in the literature.  To identify the stellar
counterpart of the host galaxies, we have carried out a near-infrared
imaging survey of these fields using PANIC (Martini \etal\ 2004) and
the $H$ filter on the Magellan Baade telescope on Las Campanas, Chile.
In addition, we have obtained and analyzed late-time deep optical
images available for some of these GRB fields from either our own
observations using the Advance Camera for Surveys (ACS; Ford \etal\
1998) or unpublished data found in the Hubble Space Telescope (HST)
data archive\footnote{http://archive.stsci.edu/hst/.}.  A summary of
the fields is presented in Table 1, where we list the GRBs, their
coordinates, redshift, and inferred neutral hydrogen column density,
$N(\hI)$, in columns (1) through (5).  The corresponding references
for the redshift and $N(\hI)$ measurements are listed in column (6).

  This GRB host sample presented here is the largest sample of GRBs at
$z_{\rm GRB}>2$ for which both ISM absorption properties and
constraints on the emission properties of the host galaxies are
available.  It offers a unique opportunity to carry out a systematic
study to understand the nature of starburst galaxies hosting GRBs at
high redshift.  The redshifts of the GRBs in our sample span a range
from $z=2.04$ to $z=4.05$ (left panel of Figure 1).  The neutral
hydrogen column densities of the GRB host ISM span a range from
$\log\,N(\hI)=16.9$ to $\log\,N(\hI)=22.6$ (right panel of Figure 1).
The sample size is restricted by the amount of available observing
time.  In comparison to GRBs with known redshift or $N(\hI)$ in the
literature, we show in Figure 1 that our sample is representative of
the spectroscopically identified GRB population in the redshift and
$N(\hI)$ parameter space.  Note that three of the 15 known GRB host
absorbers do not contain high $N(\hI)$ ($\log\,N(\hI)\ge 20.3$) that
would qualify them as a damped \lya\ absorber (e.g.\ Wolfe \etal\
2005).  The observed low $N(\hI)$ indicates that the ISM in front of
these GRBs is mostly ionized.

\begin{figure*}
\begin{center}
\includegraphics[scale=0.35,angle=0]{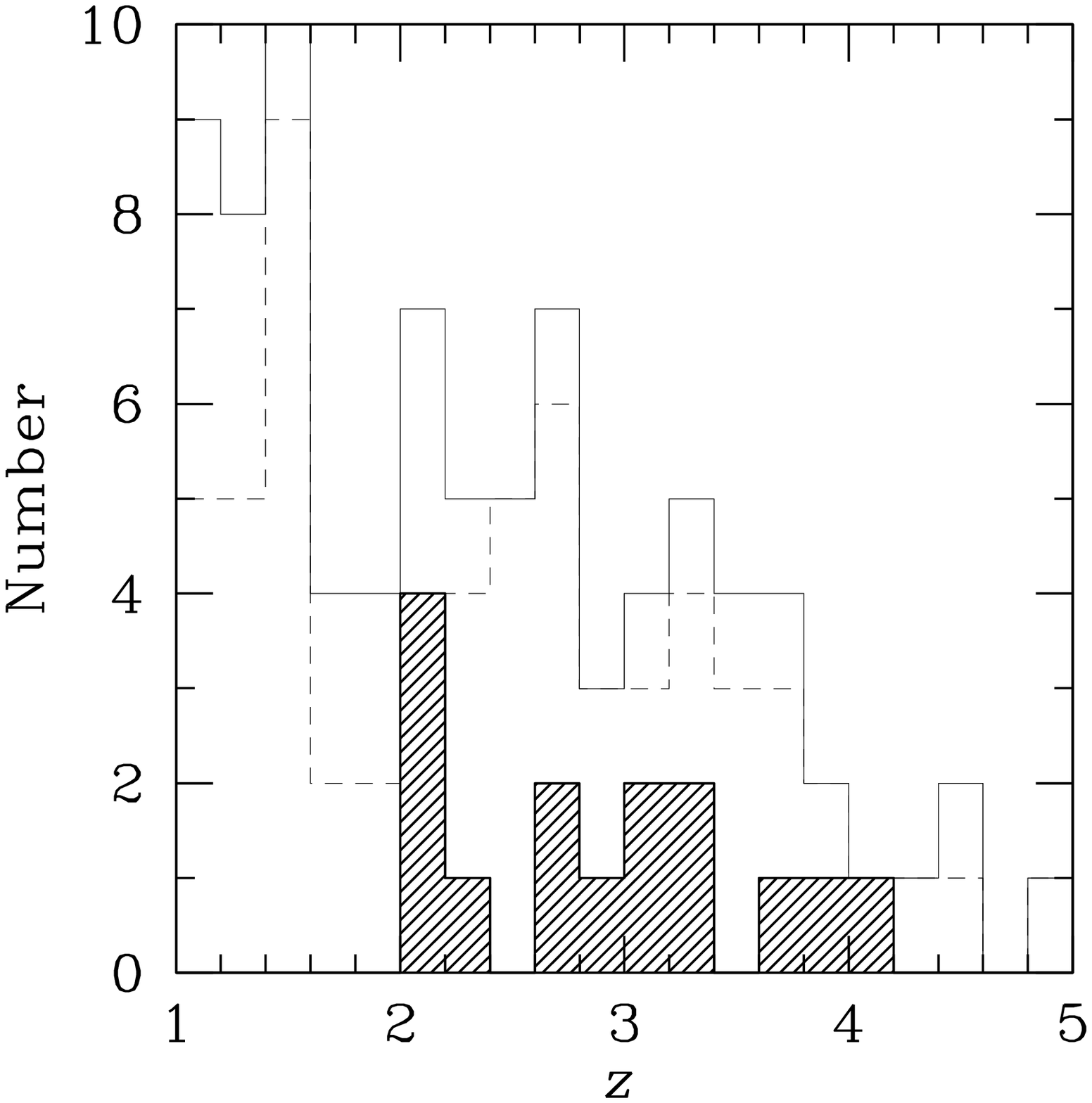}
\includegraphics[scale=0.35,angle=0]{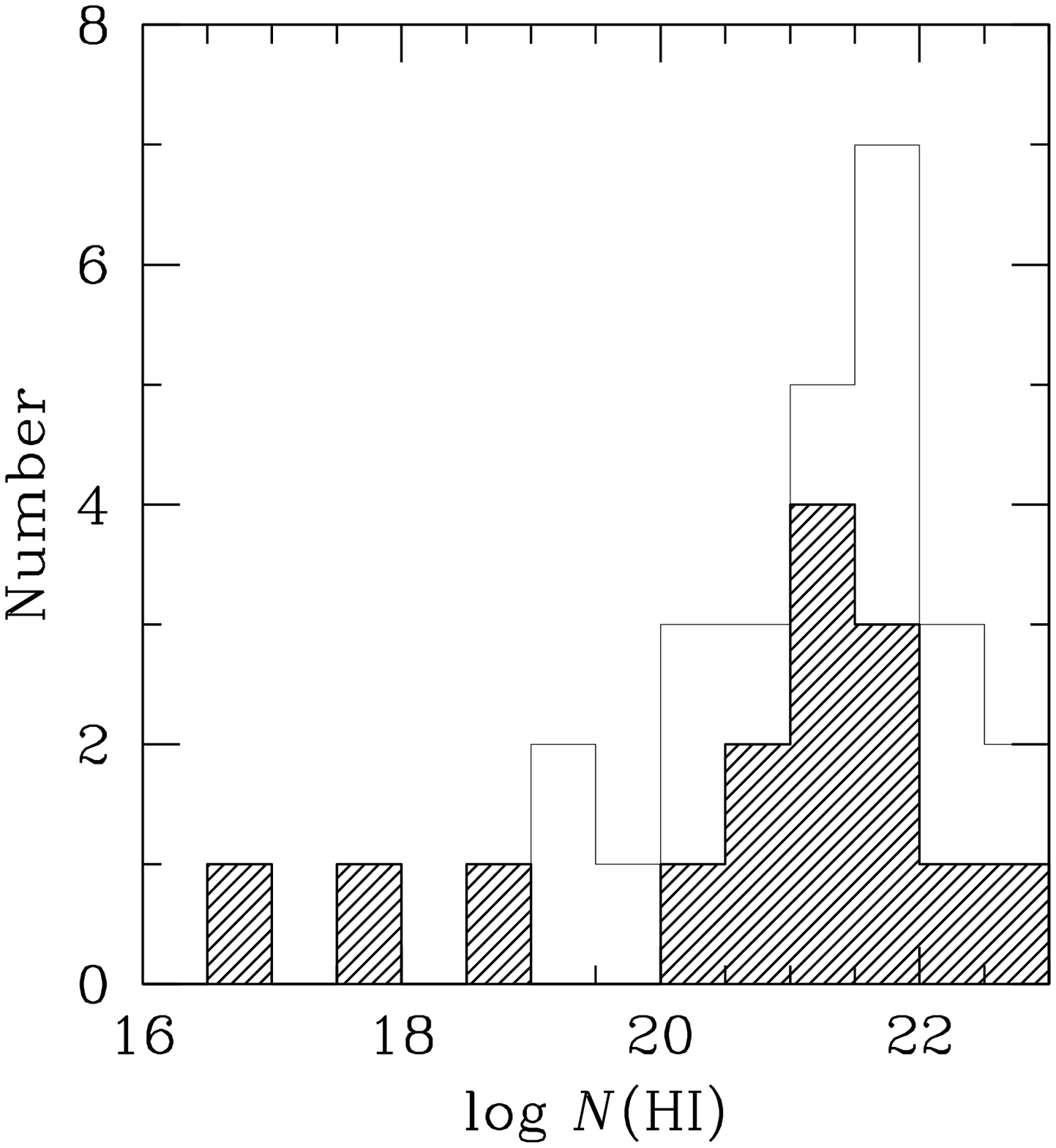}
\end{center}
\caption[]{{\it Left}: Redshift distribution of the 15 GRBs in our
sample (shaded histogram), in comparison to the distributions of all
GRBs with known redshifts (open histogram) and of those found by the
{\it Swift} Satellite (Gehrels \etal\ 2004; dashed histogram).  {\it
Right}: Neutral hydrogen column density distribution of the GRBs in
our sample in shaded histograms, in comparison to the distribution of
known host $N(\hI)$ for GRBs found prior to July 2007 (see Chen \etal\
2007a).}
\end{figure*}

\section{IMAGING OBSERVATIONS AND DATA ANALYSIS}

  To constrain the star formation and/or stellar population of the
host galaxies, we have carried out an optical and near-infrared
imaging survey of nine GRB fields that have not been studied before.
We have also obtained new near-infrared images of the fields around
GRB\,011211 (previously studied by Jakobsson \etal\ 2003) and
GRB\,030323 (previously studied by Vreeswijk \etal\ 2004), and
analyzed additional HST images of the field around GRB\,030323 that
were not included in Vreeswijk \etal\ (2004).  At $z<3$, near-infrared
images offer valuable constraints for the intrinsic luminosity of the
host galaxies at rest-frame optical wavelengths, while optical images
provide constraints for their rest-frame UV luminosities.  Here we
describe relevant imaging observations and data processing.
 
\subsection{Optical Images from the Hubble Space Telescope}

  High spatial resolution and high sensitivity optical images obtained
using the Wide Field Channel (WFC) of Advance Camera for Surveys (ACS)
on board the Hubble Space Telescope (HST) are available for
GRBs\,030323, 050820A, and 060206.  The field surrounding GRB\,030323
was observed under program ID 9405 (PI: Fruchter) using ACS/WFC and
the F606W filter during July 2003 and December 2003.  The observations
were carried out in a sequence of four exposures of between 480 and
522 s each.  The images were retrieved from the HST data archive.  The
field surrounding GRB\,050820A was observed under program ID 10551
(PI: Kulkarni) using ACS/WFC and the F625W, F775W, and F850LP filters
during September 2005 and June 2006.  The observations were carried
out in a sequence of two to four exposures of between 400 and 807 s
each.  The images were retrieved from the HST data archive.  The field
surrounding GRB\,060206 was observed under our own program (PID=10817;
PI: Chen) using ACS/WFC and the F814W filter during November and
December 2006.  The observations were carried out in the standard
``ACS-WFC-DITHER-BOX'' pattern of exposures between 1215 and 1256 s
each.

  Individual exposures were reduced using standard pipeline
techniques, corrected for geometric distortion using drizzle,
registered to a common origin, filtered for deviant pixels based on a
$5\,\sigma$ rejection criterion, and combined to form a final stacked
image.  A summary of the optical imaging observations is presented in
columns (7) through (10) of Table 1, which lists for each field the
instrument and filter used, total exposure time, and the full width at
half maximum (FWHM) of the median point spread function (PSF) as
determined from point sources.  We note that in addition to the GRBs
listed in Table 1 we will include in the following analysis previous
HST imaging observations of GRB\,000926 ($z_{\rm GRB}=2.038$) and
GRB\,021004 ($z_{\rm GRB}=2.323$) published by Fynbo \etal\ (2002),
Castro \etal\ (2003) and Fynbo \etal\ (2005), respectively.

\subsection{Optical and Near-infrared Images from the Magellan and Keck Telescopes}

  Optical $i'$ images of GRB\,050730 were obtained using MagIC on the
Magellan Baade telescope in June 2008.  The observations were carried
out in one set of three exposures, 900 s in duration.  Dither offsets
of 15\arcsec\ were applied between exposures.  The sky condition was
photometric with a mean seeing of 0.6\arcsec.  Individual exposures
were first corrected for pixel-to-pixel variation using a flat-field
image formed by median filtering sky images obtained during even
twilight.  Fringes are not apparent in the MagIC-$i'$ images.
Next,the processed individual images obtained on the same night were
registered to a common origin, filtered for deviant pixels based on a
$5\,\sigma$ rejection criterion and a bad pixel mask formed using the
flat-field frames, and stacked according to a weighting factor that is
proportional to the inverse of the sky variance.  The photometric zero
points were determined using five SDSS southern standard stars (Smith
\etal\ 2002) observed during the night.

  Optical $R_c$ and $g$ images of GRBs\,050401, 050730, 050820A,
050922C, and 060607 were obtained using the Low Resolution Imaging
Spectrometer (LRIS; Oke \etal\ 1995) on the the Keck I telescope in
May and July of 2006.  Integration times varied from source to source
and are given in Table 1, but generally consisted of dithered
exposures of about 600~s each and total integration times of 40 to 65
minutes per field.  Individual exposures were reduced and combined
using standard techniques.  Photometric calibration was performed
using a series of exposures on the Landolt field Markarian A at
three different elevations during the July run and a single visit to
the PG\,2213 field during the May run.  The observation conditions
were photometric.  Astrometry was performed using a large sample of
USNO B1.0 standard stars in each field.

  Near-infrared $H$ images of GRBs\,011211, 030323, 050401, 050820A,
and 050922c were obtained using PANIC on the Magellan Baade telescope
in February 2004, May 2006, and August 2007.  The observations were
carried out in five or nine sets of three to four exposures, 45 to 60
s in duration.  Dither offsets of eight to 15 arcsec were applied
between different sets of exposures in a slanted square pattern.

  Individual exposures were first corrected for pixel-to-pixel
variation using a flat-field image formed by median filtering all the
images obtained on the same night.  Next, we corrected for geometric
distortion in individual flat-field images using the IRAF {\it geomap}
task, according to a 2D distortion map provided by the PANIC
instrument team.  Next,the processed individual images obtained on the
same night were registered to a common origin, filtered for deviant
pixels based on a $5\,\sigma$ rejection criterion and a bad pixel mask
formed using the flat-field frames, and stacked according to a
weighting factor that is proportional to the inverse of the sky
variance.  Images obtained during non-photometric nights were scaled
to match the fluxes of common objects observed during photometric
conditions.  Individual stacked images from different nights were
combined to form a final image of each field.  The photometric zero
points were determined using three to five IR standard stars (Persson
\etal\ 1998) observed under photometric conditions.  A summary of the
near-infrared imaging observations is presented in Table 1.

\subsection{Astrometry}

  An accurate astrometric solution is necessary for the images
obtained at late times, in order to correctly identify the host
galaxies of the GRBs.  To obtain an accurate astrometric solution for
each final stacked image, we first calibrate the astrometry using
$\aapl\, 12$ USNO stars with a low-order polynomial fit.  Next, we
refine the astrometric solution using $\aapg 2$ 2MASS stars in the
image by adjusting field offsets and rotation.  We find that the final
astrometric solution is accurate to am r.m.s.\ of 0.2\arcsec.

\section{DESCRIPTION OF INDIVIDUAL FIELDS}

  Including previously published images around GRBs 000301C, 000926
and 021004, we have now collected deep optical and/or near-infrared
images surrounding 15 GRBs at $z>2$.  All of these GRBs have
early-time afterglow spectra available that allow us to determine both
the gas properties of the GRB host galaxies and the line-of-sight
properties of absorbers foreground to the hosts.  We will show that
known line-of-sight properties from afterglow spectra are important
for accurately identifying the GRB host galaxies (e.g.\ GRB\,030429,
GRB\,060206, GRB\,070721B, and see Pollack \etal\ 2008 for a rich
galaxy field around GRB\,060418).  Here we describe contraints on the
emission properties of individual host galaxies from available imaging
data, together with a brief description of known absorption-line
properties from afterglow spectroscopy.  A summary of the emission and
absorption properties of these $z>2$ GRB host galaxies is presented in
Table 3.

\subsection{GRB\,011211 at $z=2.142$}

\begin{figure}
\begin{center}
\includegraphics[scale=0.32,angle=0]{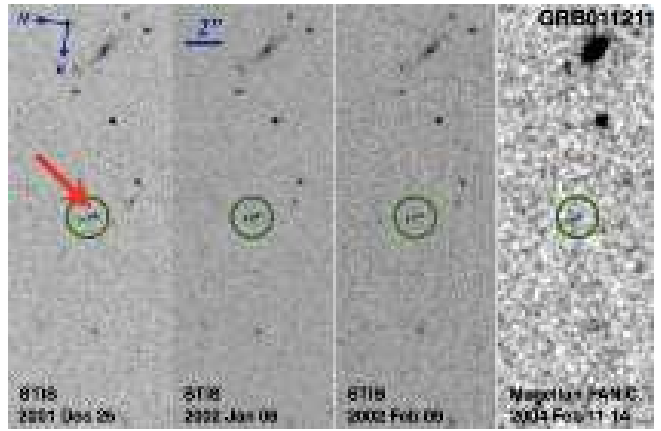}
\end{center}
\caption[]{Registered optical (left three panels) and near-infrared
$H$ (right panel) images of the field around GRB\,011211 at $z_{\rm
GRB}=2.142$ after the afterglow had faded.  The epoch during which the
images were taken is indicated at the bottom of each panel.  The host
appears to consist of three compact regions, all of which are
confirmed to be at the host redshift based on the presence of \lya\
emission in ground-based narrow-band imaging follow-up (Fynbo \etal\
2003a).  The GRB is found by Jakobsson \etal\ (2003) to originate in
the fainter region, southwest of the center blob, as indicated by the
fading optical transient (arrow in the left panel).  The mean FWHM of
the PSF in the $H$ image was found to be $\approx 0.37\arcsec$ based
on an average of 10 stars across the PANIC field.  Two of the compact
regions are detected in the $H$-band image, but the region directly
associated with the burst does not exhibit observable flux.  We
measure an $H$-band magnitude of $AB(H)=25.0\pm 0.3$ over a 1\arcsec\
diameter aperture for the host galaxy.}
\end{figure}

  This burst was detected by BeppoSAX and the optical transient was
found $\approx 10$ hours later with $R=19$ (Grav \etal\ 2001).
Low-resolution (${\rm FWHM}\equiv\delta\,v \approx 680$ \kms) optical
spectra of the afterglow were taken by different groups, which
revealed multiple absorption features indicating a source redshift of
$z=2.142$ (Fruchter \etal\ 2001; Gladders \etal\ 2001).  Analyzing
existing afterglow spectra obtained using FORS2 on the VLT telescopes,
Vreeswijk \etal\ (2006) reported a total neutral hydrogen column
density of $\log\,N(\hI)=20.4\pm 0.2$.  In addition, the authors
applied a curve-of-growth (COG) analysis over a series of absorption
features found in the afterglow spectra and derived $[{\rm Si}/{\rm
H}]=-0.9_{-0.4}^{+0.6}$ and $[{\rm Fe}/{\rm H}]=-1.3\pm
0.3$\footnote{Chemical abundances are measured relative to solar
values and defined as $[{\rm M}/{\rm H}]\equiv\log ({\rm M}/{\rm H}) -
\log ({\rm M}/{\rm H})_\odot$.}.  The large $N(\hI)$ indicates that
even the weakest absorption features detected in the afterglow
spectrum may be saturated due to the low resolution and low $S/N$ of
the data.  Here we consider the abundance measurements based on the
COG analysis lower limits to the intrisic ISM metallicity and adopt
$[{\rm Si}/{\rm H}]> -1.3$ for the ISM of the host galaxy.  However,
we note that the absence of relatively weak transitions such as Si\,II
$\lambda\,1808$ in the afterglow spectrum indicates $[{\rm Si}/{\rm
H}]< -0.7$.  Finally, this line of sight exhibits no Mg\,II absorbers
of rest-frame absorption equivalent width $W(2796)>1$ \AA\ between
$z=0.359$ and $z=2$ (Prochter \etal\ 2006).

  Imaging follow-up of the field around GRB\,011211 was carried out
during four different epochs with STIS and the clear filter on board
HST.  The imaging data were reduced and analyzed by Jakobsson \etal\
(2003), who reported the host galaxy has an $R$-band magnitude of
$AB(R)=25.15\pm 0.11$ over a 1\arcsec-diameter aperture.  In addition,
\lya\ emission was detected in a ground-based narrow-band imaging
survey by Fynbo \etal\ (2003a), who measured a total flux of
$f(\lya)=(2.8\pm 0.8)\times 10^{-17}$ erg s$^{-1}$ cm$^{-2}$ and
derived an SFR of $0.8\pm 0.2$ M$_\odot$ yr$^{-1}$.

  We have observed this field using PANIC on Magellan in February
2004, and obtained a total integration of 216 minutes.  The mean FWHM
of the PSF was found to be $\approx 0.37\arcsec$ based on an average
of 10 stars across the PANIC field.  The final stacked image is
presented in Figure 2, together with HST STIS images obtained roughly
14, 26, and 32 days after the initial burst.  The optical and $H$
images are registered to a common origin.  The host galaxy is marked
by a circle of 1\arcsec\ radius.  The host appears to consist of three
compact (presumably star-forming) regions in the STIS images, all of
which are confirmed to be at the host redshift based on the presence
of \lya\ emission in the ground-based narrow-band images presented in
Fynbo \etal\ (2003a).  The GRB is found to originate in the faintest of
the three blobs, southeast of the center one.  Two of the compact
regions are detected in the $H$-band image, but the region directly
associated with the burst does not exhibit any detectable flux.  We measure
an $H$-band magnitude of $AB(H)=25.0\pm 0.3$ over a 1\arcsec\ diameter
aperture for the host galaxy.  

  At $z=2.142$, the observed $R$-band magnitude corresponds to a
rest-frame absolute magnitude at 2000 \AA\ of
$M(2000)-5\,\log\,h=-19.2\pm 0.1$.  The observed $H$-band magnitude
allows us to derive a rest-frame absolute $B$-band magnitude of
$M_{AB}(B) - 5\,\log\,h=-19.1\pm 0.3$ for the GRB host galaxy.

\subsection{GRB\,020124 at $z_{\rm GRB}=3.198$}

  This burst was detected by HETE (Ricker \etal\ 2002) and the optical
transient was found $\approx 22$ hours later with $R\sim 18.5$ (Price
\etal\ 2002).  Low-resolution ($\delta\,v \approx 680$ \kms) optical
spectra of the afterglow were obtained using FORS1 on the VLT Melipal
telescope, which revealed multiple absorption features indicating a
source redshift of $z=3.198$ (Hjorth \etal\ 2003).  The total neutral
hydrogen column density derived based on the observed \lya\ absorption
line is $\log\,N(\hI)=21.7\pm 0.2$ (Hjorth \etal\ 2003).  The low
resolution and low $S/N$ ($\approx\, 3-4$) of the data did not allow
an accurate measurement of the chemical content in the host ISM.  No
information is available for the line-of-sight properties of
additional intervening absorbers.

\begin{figure}
\begin{center}
\includegraphics[scale=0.3,angle=270]{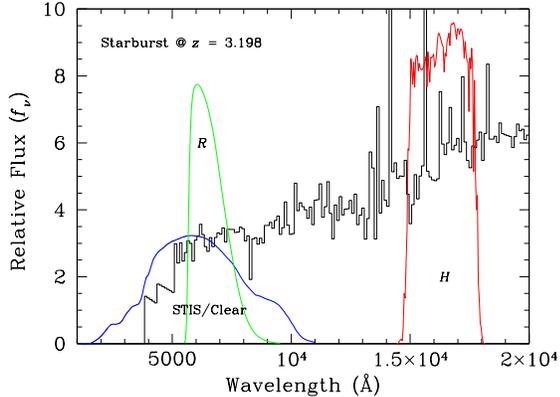}
\end{center}
\caption[]{Schematic diagram to illustrate the photometric zero point
offset neccessary to be included for sources at $z=3.2$, from a clear
bandpass that covers a spectral range over $\lambda=2000-10,000$ \AA\
to a typical Johnson $R$ filter.  The historgram represents a typical
starburst spectral template with additional absorption at
$\lambda_{\rm rest} \le 1215$ \AA\ due to the IGM \lya\ forest and the
ISM internal to the host.}
\end{figure}

  Imaging follow-up of the field around GRB\,020124 was carried out
roughly 18 and 21 days after the burst with STIS and the clear filter
on board HST.  The imaging data were reduced and analyzed by Berger
\etal\ (2002).  While the OT was detected in the first epoch image, no
detectable flux was found at the OT position in the second epoch image.
Berger \etal\ (2002) placed an upper limit for the $R$-band magnitude
of the host galaxy at $R>29.5$.  We have retrieved the imaging data
from the HST archive and determined a 5-$\sigma$ limit of the second
epoch image at $AB({\rm clear})=29.4$ over a 0.5\arcsec\ diameter
aperture.  To derive the corresponding detection limit, we take into
account the bandpass difference between STIS/clear and $R$ and the IGM
opacity that reduces most of the light at $\lambda_{\rm obs}\le 4000$
\AA\ for sources at $z=3.198$ (Figure 3).  We apply a 0.4 mag offset
in the photometric zero point and conclude that the host galaxy is
fainter than $AB(R)=29$ at the 5-$\sigma$ level of significance.

  We have also observed this field using PANIC on Magellan in February
2004, and obtained a total integration of 231 minutes.  The mean FWHM
of the PSF was found to be $\approx 0.4\arcsec$.  The final stacked
image is presented in Figure 4, together with available HST STIS
images.  The optical and $H$ images are registered to a common origin.
The $H$-band image has been smoothed using a Gaussian kernel of ${\rm
FWHM}=0.4\arcsec$.  The position of the GRB is marked by a circle of
0.5\arcsec radius.  No detectable flux is found at the location of
the GRB, but some emission features are observed with $AB(H)\approx
26.4$ at $\Delta\,\theta=0.4\arcsec$ angular distance away.  We place
a 2-$\sigma$ limit in the observed $H$-band magnitude of $AB(H)=26.1$
over a 0.5\arcsec\ diameter aperture for the host galaxy.  At
$z=3.198$, the observed $H$-band magnitude limit allows us to derive a
limiting rest-frame absolute magnitude of $M_{AB}(3960) -
5\,\log\,h>-18.8$ for the GRB host galaxy.

\begin{figure}
\begin{center}
\includegraphics[scale=0.32,angle=0]{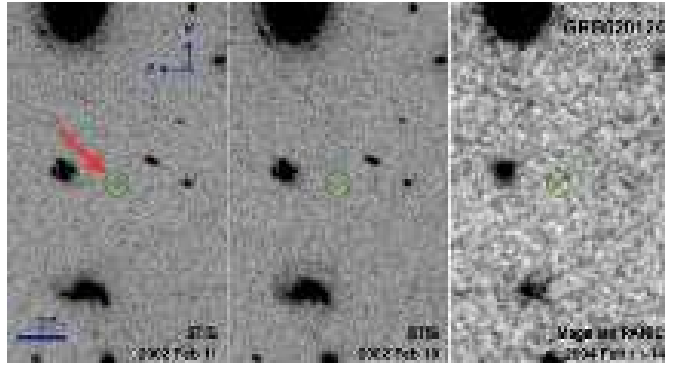}
\end{center}
\caption[]{Registered optical (left two panels) and near-infrared $H$
(right panel) images of the field around GRB\,020124 at $z_{\rm
GRB}=3.198$.  The epoch during which the images were taken is
indicated at the bottom of each panel.  The $H$-band image has been
smoothed using a Gaussian kernel of ${\rm FWHM}=0.4\arcsec$, which is
roughly the size of the PSF.  While the OT is still visible in the
first-epoch image (left panel), the host is not detected in either of
the two late-time images.  We place a 2-$\sigma$ $H$-band limiting
magnitude of $AB(H)=26.1$ over a 0.5\arcsec\ diameter aperture for the
host galaxy.}
\end{figure}

\subsection{GRB\,030323 at $z_{\rm GRB}=3.372$}

  This burst was detected by HETE-2 (Graziani \etal\ 2003) and the
optical transient was found $\approx 7.6$ hours later with $R=18.7$
(Gilmore \etal\ 2003a).  Moderate-resolution ($\delta\,v \approx 150$
\kms) optical spectra of the afterglow, covering $\lambda=4560-7310$
\AA, were taken using FORS2 on the VLT Yepun telescope by Vreeswijk
\etal\ (2004), revealing multiple absorption features from ions in
both ground states and excited states that are consistent with a
source redshift of $z=3.372$.  These authors estimated the total
neutral hydrogen column density of $\log\,N(\hI)=21.90\pm 0.07$ in the
host ISM and chemical abudances of $[{\rm S}/{\rm H}]=-1.26\pm 0.2$
and $[{\rm Fe}/{\rm H}]=-1.47\pm 0.11$.  The observed large column
densities of various ions suggest that these reported values represent
only lower limits to the intrinsic abundances of these ions.  We
therefore adopt $[{\rm S}/{\rm H}]> -1.26$ for the ISM of the host
galaxy.  This line of sight exhibits no strong Mg\,II absorbers at
$z=0.824-1.646$ (Prochter \etal\ 2006).

  Imaging follow-up of the field around GRB\,030323 was carried out
during two different epochs with ACS and the F606W filter on board
HST.  The first epoch imaging data with a total exposure time of 1920
s were reduced and analyzed by Vreeswijk \etal\ (2004), who identified
the host galaxy at $\Delta\,\theta=0.14\arcsec$ from the position of
the OT and measured $AB({\rm F606W})=28.0\pm 0.3$ over a 0.3\arcsec\ 
diameter aperture.  Additional imaging data were obtained five month
later with HST using the same instrument setup.  A stack of all
available imaging data from the HST data archive shows a clear
detection of the host galaxy at $\Delta\,\theta=0.22\arcsec$ from the
position of the OT.  This position is consistent with the position of
Vreeswijk \etal\ to within the astrometric uncertainties.  After
correcting for the Galactic extinction ($E(B-V)=0.049$ according to
Schlegel \etal\ 1998), we measure a total flux of $AB({\rm
F606W})=27.4\pm 0.1$ over a 0.5\arcsec\ diameter aperture.

  We have also observed this field using PANIC on Magellan in February
2004, and obtained a total integration of 240 minutes.  The mean FWHM
of the PSF was found to be $\approx 0.4\arcsec$.  The final stacked
image is presented in Figure 5, together with a stack of available
ACS/F606W images.  The optical and $H$ images are registered to a
common origin.  The host galaxy is marked by a circle of 0.5\arcsec\ 
radius.  The host appears to be extended in the ACS image, but not
detected in the stacked $H$ image.  We place a 2-$\sigma$ limit in the
observed $H$-band magnitude of $AB(H)=26.2$ over a 0.5\arcsec\ diameter
aperture for the host galaxy.  

\begin{figure}
\begin{center} 
\includegraphics[scale=0.38,angle=0]{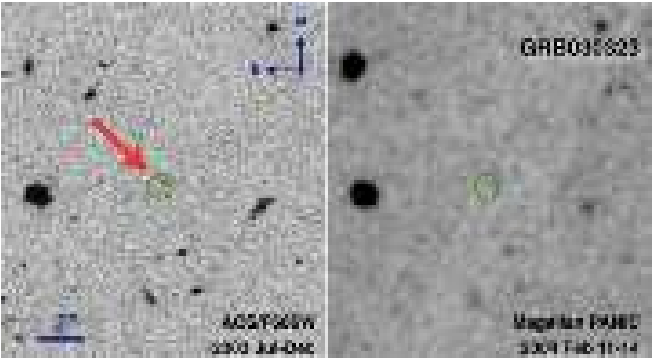}
\end{center} 
\caption[]{Registered optical (left panel) and near-infrared $H$
(right panel) images of the field around GRB\,030323 at $z_{\rm
GRB}=3.372$.  The epoch during which the images were taken is
indicated at the bottom of each panel.  The host galaxy is identified
at $\Delta\,\theta=0.22\arcsec$ from the position of the OT with
$AB({\rm F606W})=27.4\pm 0.1$ over a 0.5\arcsec\ diameter aperture.
The $H$-band image has been smoothed using a Gaussian kernel of ${\rm
FWHM}=0.4\arcsec$, which is roughly the size of the PSF.  The host is
not detected in the $H$ image.  We place a 2-$\sigma$ $H$-band
limiting magnitude of $AB(H)=26.2$ over a 0.5\arcsec\ diameter aperture
for the host galaxy.}
\clearpage
\end{figure}

  At $z=3.372$, the observed F606W magnitude corresponds to
$M_{AB}(1400) - 5\,\log\,h=-17.6\pm 0.1$ for the GRB host galaxy.  The
observed $H$-band magnitude limit allows us to derive a limiting
rest-frame absolute magnitude of $M_{AB}(3800) - 5\,\log\,h>-18.8$ for
the GRB host galaxy.

\subsection{GRB\,030429 at $z_{\rm GRB}=2.658$}

  This burst was detected by HETE-2 (Doty \etal\ 2003) and the optical
transient was found $\approx 3.5$ hours later (Gilmore \etal\ 2003b)
at $\Delta\,\theta\approx 1.2\arcsec$ southeast of an extended source
of $R\approx 24$ (Fynbo \etal\ 2003b).  Low-resolution ($\delta\,v
\approx 680$ \kms) optical spectra of the afterglow and the extended
source were obtained using FORS1 on the VLT Melipal telescope.  The
spectrum of the afterglow exhibits multiple absorption features that
are consistent with $z=2.658$ (Jakobsson \etal\ 2004).  In contrast,
the spectrum of the extended source exhibits a single emission feature
at $\lambda=6858$ \AA.  Jakobsson \etal\ (2004) identified this
emission as [O\,II] at $z=0.841$, which is supported by the presence
of an absorption feature in the afterglow spectrum if interpreted as a
Mg\,II absorber at $z=0.841$.  The GRB host galaxy therefore remains
unidentified.  These authors also reported a total neutral hydrogen
column density of $\log\,N(\hI)=21.6\pm 0.2$ based on the observed
\lya\ absorption feature.  The low resolution of the spectrum did not
allow an accurate measurement of the chemical content in the host ISM.

\begin{figure}
\begin{center}
\includegraphics[scale=0.38,angle=0]{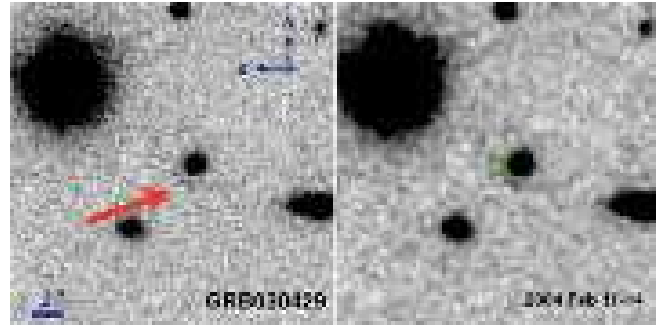}
\end{center} 
\caption[]{Near-infrared $H$ image of the field around GRB\,030349 at
$z_{\rm GRB}=2.658$.  The mean FWHM of the PSF is 0.35\arcsec.  A
smooth version of the $H$-band image is presented in the right panel.
No detectable flux is found at the immediate location of the GRB, but
extended emission features (pointed by an arrow in the left panel) are
clearly visible at $\approx 1\arcsec$ southwest of the OT, or
1.3\arcsec\ south of the galaxy at $z=0.841$.  We measure
$AB(H)=20.57\pm 0.05$ for the foreground galaxy, and $AB(H)=24.4\pm
0.1$ for the host candidate.}
\clearpage
\end{figure}

  We have observed the field around GRB\,030429 using PANIC on
Magellan in February 2004, and obtained a total integration of 129
minutes.  The mean FWHM of the PSF was found to be $\approx
0.35\arcsec$.  The final stacked image is presented in Figure 6,
together with a smoothed version with a Gaussian kernel of ${\rm
FWHM}=0.4\arcsec$.  The position of the GRB is marked by a circle of
0.5\arcsec radius.  No detectable flux is found at the immediate
location of the GRB, but additional emission features are present at
$\approx 1\arcsec$ southwest of the OT, or 1.3\arcsec\ south of the
galaxy at $z=0.841$.  We measure $AB(H)=20.57\pm 0.05$ for the
foreground galaxy, and $AB(H)=24.4\pm 0.1$ for the host candidate.

  The emission morphology resembles those found for the hosts of
GRB\,000926 and GRB\,011211, with the OT originating in the faintest
of the three emission blobs.  The close proximity to the foreground
galaxy suggests that the host may be gravitationally lensed by this
foreground object, although the large angular separation implies that
the foreground galaxy would have to be $\approx\,4\times$ as massive
as the Milky Way (see also Jakobsson \etal\ 2004).  Additional images
at optical wavelengths are necessary to investigate this lensing
hypothesis further.

  If the extended feature is associated with the host galaxy, then the
OT would be at a projected distance of $\rho=5.4\ h^{-1}$ kpc from the
center of the galaxy.  In the following discussion, we consider the
observed brightness as an upper limit to the brightness of the GRB
host and derive a rest-frame $B$-band magnitude of $M_{AB}(B) -
5\,\log\,h>-20.1$ for the GRB host galaxy.

\subsection{GRB\,050401 at $z_{\rm GRB}=2.899$}

  This burst was detected by {\it Swift} (Barbier \etal\ 2005) and
prompt localization of the source was reported by Angelini \etal\
(2005) from the X-ray afterglow.  The optical transient was found
$\approx 1$ hour later with $R=20.3$ (McNaught \& Price 2005).
Low-resolution ($\delta\,v \approx 680$ \kms) optical spectra of the
afterglow were taken using FORS2 on the VLT Antu telescope (Watson
\etal\ 2006), revealing multiple absorption features from ions in both
ground states and excited states that are consistent with a source
redshift of $z=2.899$.  Watson \etal\ (2006) estimated the total
neutral hydrogen column density of $\log\,N(\hI)=22.6\pm 0.3$ in the
host ISM and metallicity of $[{\rm Zn}/{\rm H}]=-1.0\pm 0.4$.
Similiar to the other GRB hosts where only low-resolution afterglow
spectra are available, the observed large column densities of various
ions suggest that these reported values represent only lower limits to
the intrinsic abundances of these ions.  We therefore adopt $[{\rm
Zn}/{\rm H}]> -1.0$ for the ISM of the host galaxy.  An additional set
of ionic transitions (such as Al\,II and Fe\,II) was also reported at
$z=2.5$ by Watson \etal\ (2006).

  We have observed the field around GRB\,050401 using PANIC on
Magellan in May 2006, and obtained a total integration of 105 minutes.
The mean FWHM of the PSF was found to be $\approx 0.6\arcsec$.  No
detectable flux is seen at the immediate location of the GRB.  We
place a 2-$\sigma$ limit in the observed $H$-band magnitude of
$AB(H)=25.1$ over a 1\arcsec\ diameter aperture for the host galaxy.
At $z=2.899$, the observed $H$-band magnitude limit allows us to
derive a limiting rest-frame absolute magnitude of $M_{AB}(B) -
5\,\log\,h>-19.6$ for the GRB host galaxy.

  Optical images of the field around GRB\,050401 have also been
obtained using LRIS and the $g$ and $R_c$ filters on the Keck I
telescope in May 2006.  The mean FWHM of the PSF was found to be
$\approx 1.1\arcsec$ in the combined $g$ image and $\approx
1.0\arcsec$ in the combined $R_c$ image.  The stacked images are
presented in Figure 7, which have been smoothed with a Gaussian kernel
of ${\rm FWHM}=1\arcsec$.  The position of the GRB is marked by a
circle of 2\arcsec\ radius.  At the location of the afterglow reported
by (McNaught \& Price 2005), we detect faint emission in both images
at roughly 2-$\sigma$ significance level.  We estimate a total
brightness of $AB(g)=27.5\pm 0.5$ and $AB(R_c)=27.3\pm 0.4$ over a
2\arcsec\ diameter aperture for the object.  Due to the presence of a
strong absorber at $z=2.5$, the identification of the observed faint
emission is uncertain.  We note, however, that at $z=2.899$ the IGM
\lya\ forest is expected to absorb a large fraction of flux in the $g$
band, resulting in $g-R_c\approx 0.2-0.4$ mag for a flat spectrum
source.  Additional HST imaging data are necessary to confirm the host
identification.  At $z=2.899$, the observed $R_c$-band magnitude would
imply a rest-frame absolute magnitude of $M_{AB}(1600) -
5\,\log\,h=-17.4\pm 0.4$ for the candidate host galaxy.

\begin{figure}
\begin{center} 
\includegraphics[scale=0.31,angle=0]{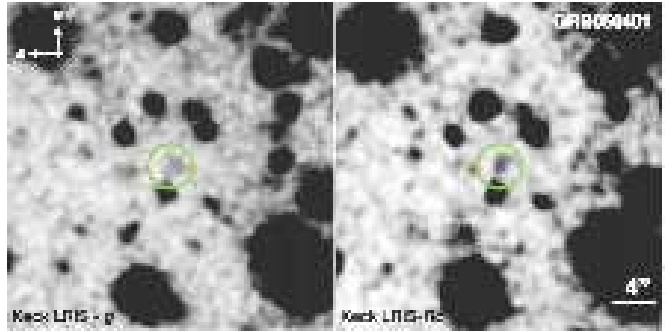}
\end{center}
\caption[]{Optical $g$ (left panel) and $R_c$ (right panel) images of
the field around GRB\,050401 at $z_{\rm GRB}=2.899$.  The mean seeing
is 1.1\arcsec\ and 1.0\arcsec\ in $g$ and $R_c$, respectively. At the
location of the afterglow reported by (McNaught \& Price 2005), we
detect faint emission in both images at roughly 2-$\sigma$
significance level.  The circle indicates a 2\arcsec\ angular radius
around the afterglow position.}
\end{figure}

\subsection{GRB\,050730 at $z_{\rm GRB}=3.968$}

  This burst was detected by {\it Swift} (Holland \etal\ 2005).  An
optical transient was found promptly using the Ultraviolet-Optical
Telescope (UVOT) on board Swift with $V = 17.6$ about three minutes
after the burst trigger (Holland \etal\ 2005).  We obtained an echelle
spectrum of the afterglow, using the MIKE spectrograph (Bernstein
\etal\ 2003) on the Magellan Clay telescope, four hours after the
initial trigger.  Descriptions of the data were presented in Chen
\etal\ (2005) and Prochaska \etal\ (2007b).  The spectrum covers a
full spectral range from 3300 \AA\ through 9400 \AA\ with a spectral
resolution of $\delta\,v \approx 10$ \kms\ at wavelength
$\lambda=4500$ \AA\ and $\delta\,v \approx 12$ \kms\ at $\lambda=8000$
\AA.  The host of the GRB exhibits a strong damped \lya\ absorption
feature with $\log\,N(\hI)=22.15\pm 0.05$, and metallicity $[{\rm
S}/{\rm H}]=-2.26\pm 0.1$ and $[{\rm S}/{\rm Fe}]> +0.24\pm 0.11$ (see
also Starling \etal\ 2005; D'Elia \etal\ 2007).  The low metallicity
and low $\alpha$-element enhancement implies a nearly dust free
medium.  No trace of H$_2$ is found despite the large $N(\hI)$.  The
high-resolution, high $S/N$ echelle data allow us to place a sensitive
limit on the molecular fraction of the host ISM at $f_{\rm H_2}\equiv
2N({\rm H_2})/[N(\hI)+2\,N({\rm H_2})] < 10^{-7.1}$ (Tumlinson \etal\
2007).  Additional strong absorbers are found at $z=3.56$, $z=3.02$,
$z=2.25$, and $z=1.77$ (Chen \etal\ 2005).

  We have observed the field around GRB\,070530 using MagIC and the
$i'$ filter on Magellan in June 2008, and obtained a total integration
of 45 minutes.  The mean FWHM of the PSF was found to be $\approx
0.6\arcsec$.  The final stacked image is presented in Figure 8, which
has been smoothed using a Gaussian kernel of ${\rm FWHM}=0.6\arcsec$.
The position of the GRB is marked by a circle of 1\arcsec\ radius.  No
detectable flux is seen at the immediate location of the GRB.  We
place a 2-$\sigma$ limit in the observed $i'$-band magnitude of
$AB(i')=26.6$ over a 1\arcsec\ diameter aperture for the host galaxy.
At $z=3.968$, observed $i'$-band magnitude limit allows us to derive a
limiting rest-frame absolute magnitude of $M_{AB}(1500) -
5\,\log\,h>-18.6$ for the GRB host galaxy.

  Optical images of the field around GRB\,050730 have also been
obtained using LRIS and the $g$ and $R_c$ filters on the Keck I
telescope in May 2006.  The mean FWHM of the PSF was found to be
$\approx 1.2\arcsec$ in the combined $g$ image and $\approx
1.0\arcsec$ in the combined $R_c$ image.  No detectable flux is
seen at the immediate location of the GRB.  We place 2-$\sigma$
limits of $AB(g)=26.6$ and $AB(R_c)=26.4$ over a 2\arcsec\ diameter
aperture for the host galaxy.  Taking into account that at $z=3.968$
the IGM \lya\ forest and the large amount of neutral gas in the host
ISM absorb a large fraction of flux in the $g$ and $R_c$ bands, we
find that the flux limits derived from the LRIS images are consistent
with the flux limit seen in the MagIC $i'$-band data.

\begin{figure}
\begin{center}
\includegraphics[scale=0.4,angle=0]{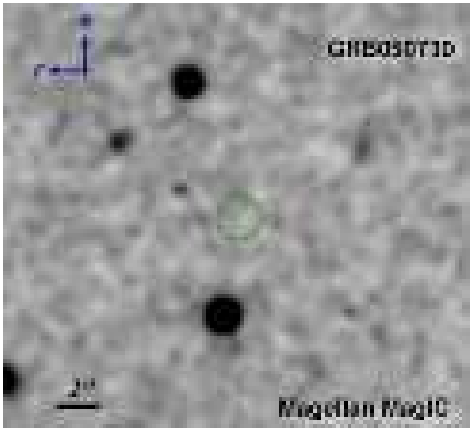}
\end{center}
\caption[]{A smoothed $i'$ image of the field around GRB\,050730 at
$z_{\rm GRB}=3.968$.  No detectable flux is seen at the location of
the OT.  We place a 2-$\sigma$ $i'$-band limiting magnitude of
$AB(H)=26.6$ over a 1\arcsec\ diameter aperture for the host galaxy.}
\end{figure}

\subsection{GRB\,050820A at $z_{\rm GRB}=2.615$}

  This burst was detected by {\it Swift} (Page \etal\ 2005a).  An
optical transient, reported less than 1 hour after the GRB, was
identified in data taken shortly after the trigger (Fox \& Cenko 2005;
Vestrand \etal\ 2006).  High-resolution ($\delta\,v \approx 7-10$
\kms) echelle spectra of the afterglow obtained shortly after the
burst are available from both our own observations using HIRES (Vogt
\etal\ 1994) on the Keck I telescope and the ESO data archive for UVES
(D'Odorico \etal\ 2000).  These two spectra together cover a full
spectral range from 3300 \AA\ through 10,000 \AA, allowing accurate
estimates of chemical abudances in the ISM of the GRB host galaxy.
Based on multiple absorption features from both ground-state and
excited-state ions, we determine a source redshift of $z=2.6147$ and a
total neutral hydrogen column density of $\log\,N(\hI)=21.0\pm 0.1$
(Prochaska \etal\ 2007b; Ledoux \etal\ 2005).  An absorption-line
analysis of various ions shows that $[{\rm S}/{\rm H}]= -0.63\pm 0.11$
and $[{\rm S}/{\rm Fe}]= +0.97\pm 0.09$ (Prochaska \etal\ 2007b), from
which we derive a dust-to-gas ratio that is comparable to what is seen
in the Small Magellanic Cloud (SMC).  Adopting the SMC dust-to-gas
ratio (Gordon \etal\ 2003), we estimate a visual extinction in the
host ISM of $A_V\approx 0.08$.  Finally, no trace of H$_2$ is found
despite the large $N(\hI)$ and moderate metallicity.  The
high-resolution, high $S/N$ echelle data allow us to place a sensitive
limit on the molecular fraction of the host ISM at $f_{\rm H_2}\equiv
2N({\rm H_2})/[N(\hI)+2\,N({\rm H_2})] < 10^{-6.5}$ (Tumlinson \etal\
2007).

  Imaging follow-up of the field around GRB\,050820A was carried out
with ACS and the F625W, F775W, F850LP filters on board HST in two
epochs, roughly 37 days, and nine months after the burst
(PID$=$10551).  We retrieved the imaging data from the HST data
archive and processed the images following the descriptions in \S\ 3.
Stacked images from the two epochs are presented in Figure 9.  The OT
was clearly detected in the first epoch image, but faded in the second
epoch image which reveals faint extended emission features of the host
galaxy.  In addition to the extended low-surface feature seen at the
position of the OT, which is identified as the host galaxy, we
identify two compact sources at $\Delta\,\theta=1.3\arcsec$ north of
the OT (Object $A$) and $0.4\arcsec$ south of the OT (Object $B$) from
the afterglow lines of sight.  We measure isophotal magnitudes of the
host and Objects $A$ and $B$.  The isophotal apertures are defined
based on the extent of objects found in a ``white light'' image, which
is a stack of the F625W, F775W, and F850LP images.  This ``white
light'' image has the optimal $S/N$ for recovering faint emission
features and allows us to determine a common aperture for every object
across different bandpasses.  We correct the observed brightness to
account for intrinsic absorption in the Milky Way ($E(B-V)=0.044$
along the line of sight according to Schlegel \etal\ 1998).  The
photometric measurements are presented in Table 2.

\begin{figure*} 
\begin{center}  
\includegraphics[scale=0.5,angle=0]{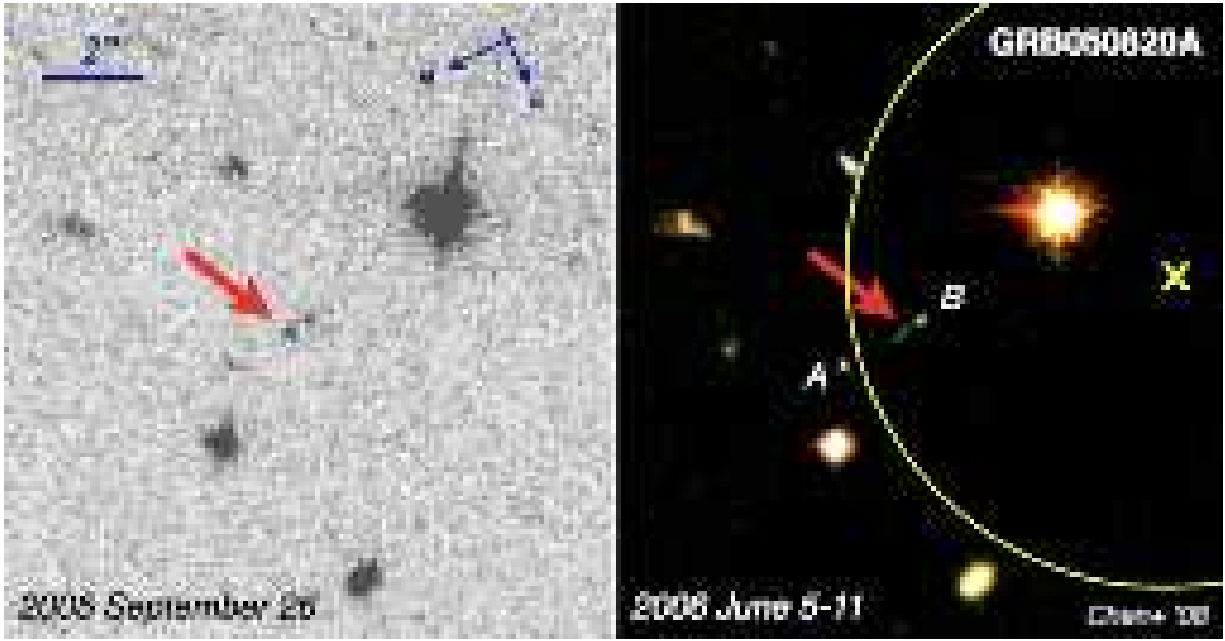}
\end{center}
\caption[]{Images of the field around GRB\,050820A at $z_{\rm
GRB}=2.6147$.  The first-epoch image (left panel), taken $\approx 37$
days after the burst with ACS and the F850LP filter, clearly shows the
OT (indicated by the arrow).  The second-epoch image (right panel),
taken nine months later, reveals extended faint blue emission of the
host galaxy.  The false-color image was formed by combining stacks
of F625W, F775W, and F850LP images.  The yellow cross and circle mark
the position of the x-ray afterglow and associated error reported by
Page \etal\ (2005b).  In addition to the host galaxy, we also point out
two compact sources $A$ and $B$ at $\Delta\,\theta=1.3\arcsec$ and
$0.4\arcsec$, respectively, from the afterglow lines of sight.}
\end{figure*}

  We have also observed this field using PANIC on Magellan in August
2007, and obtained a total integration of 141 minutes.  The mean FWHM
of the PSF was found to be $\approx 0.5\arcsec$.  The final stacked
image is presented in Figure 10, together with the second-epoch
ACS/F850LP image.  The optical and $H$ images are registered to a
common origin, and the $H$-band image has been smoothed using a
Gaussian kernel of ${\rm FWHM}=0.5\arcsec$.  The position of the GRB
is marked by a circle of 0.5\arcsec\ radius.  Extended emission at the
position of the host galaxy is detected in the $H$ image at $\approx
3\,\sigma$ level of significance.  We measure an $H$-band magnitude of
$AB(H)=25.3\pm 0.3$ within the isophotal aperture defined for the host
galaxy in the ``white light'' image described above.  

  At $z=2.6147$, the observed F775W magnitude of the host galaxy
corresponds to a rest-frame absolute magnitude at 2000 \AA\ of
$M(2000)-5\,\log\,h=-18.2\pm 0.06$.  The observed $H$-band magnitude
allows us to derive a rest-frame absolute $B$-band magnitude of
$M_{AB}(B) - 5\,\log\,h=-19.2\pm 0.3$ for the GRB host galaxy.
Adopting $A_V=0.08$ estimated above from absorption-line abundance
ratios and the SMC extinction law (Gordon \etal\ 2003), we derive
extinction corrected rest-frame absolute magnitudes of $M_{AB}(B) -
5\,\log\,h=-19.3\pm 0.3$ and $M(2000)-5\,\log\,h=-18.5\pm 0.06$.

\begin{figure*} 
\begin{center}
\includegraphics[scale=0.6,angle=0]{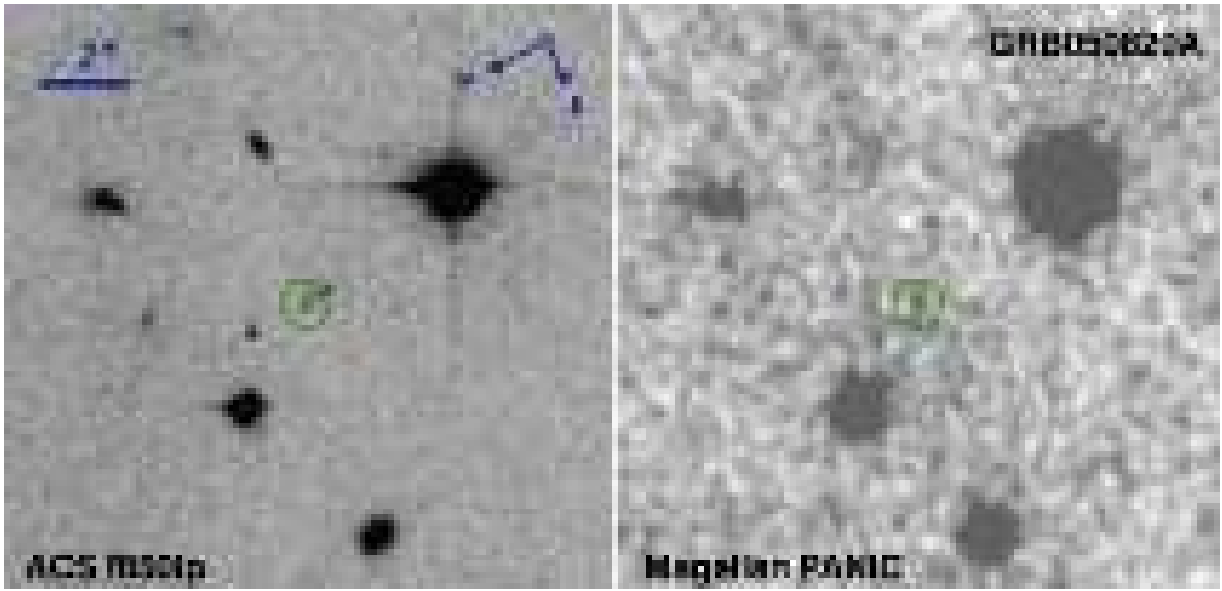}
\end{center}
\caption[]{Registered ACS F850LP image (left panel) and near-infrared
$H$ (right panel) image of the field around GRB\,050820A at $z_{\rm
GRB}=2.6147$.  The $H$-band image has been smoothed using a Gaussian
kernel of ${\rm FWHM}=0.5\arcsec$, which is roughly the size of the 
PSF.  The host is detected in the $H$ image with $AB(H)=25.3\pm 0.3$.
Neither Object $A$ or $B$ exhibits faint emission in the $H$ image.  We
place a 2-$\sigma$ limit of $AB(H)>26$ for the two sources.}
\clearpage
\end{figure*}

  Neither Object $A$ or $B$ exhibits detectable flux in the $H$ image.
We place a 2-$\sigma$ limit of $AB(H)>26$ for the two sources.  The
$H$-band photometric measurements are also presented in Table 2.  The
observed optical and near-infrared colors of Objects $A$ and $B$ are
relatively bluer than those of the GRB host galaxy and inconsistent
with the expectations for $z>2$ star-forming objects.  This suggests
that these are likely foreground galaxies.  At these small angular
separations ($\Delta\,\theta < 1.4$), objects at $z<2.6$ have
projected distances of $\rho< 8\ h^{-1}$ kpc to the afterglow line of
sight, and are expected to imprint strong Mg\,II
$\lambda\lambda\,2796, 2803$ absorption features in the afterglow
spectrum (e.g.\ Chen \& Tinker 2008).

  Incidentally, two strong Mg\,II absorbers are found at $z=0.692$ and
$z=1.430$ with $W(2796)=2.99\pm 0.03$ \AA\ and $W(2796)=1.9\pm 0.1$
\AA\ in the rest frame, respectively.  The $z=0.692$ Mg\,II absorber
exhibits a complex kinematic profile with multiple absorption
components spreading over a line-of-sight velocity interval of
$\Delta\,v\approx 500$ \kms, and a non-negligible amount of Ca$^+$
ions (see Figure 13 in Prochaska \etal\ 2007b).  The complex kinematic
profile of the absorber and the presence of Ca$^+$ together may be
explained by a sightline passing through an interacting system,
simliar to the Milky Way and the Magellanic Stream (e.g.\ Gibson
\etal\ 2000; Putman \etal\ 2003).  Attributing both Objects $A$ and
$B$ to the $z=0.692$ absorber implies a total absolute $B$-band
magnitude of $M_{AB}(B) - 5\,\log\,h=-16.27\pm 0.05$.  Follow-up
near-infrared H$\alpha$ spectroscopy is necessary to provide
conclusive identifications of these two sources.

  Optical images of the field around GRB\,050820A have also been
obtained using LRIS and the $g$ and $R_c$ filters on the Keck I
telescope in July 2006.  The mean FWHM of the PSF was found to be
$\approx 0.7\arcsec$ in the combined $g$ image and $\approx
1.9\arcsec$ in the combined $R_c$ image.  The effective seeing was
significantly compromised in the $R_c$ images due to a focus problem
with the LRIS red-side.  Faint emission is clearly detected at the
position of the afterglow, but we are unable to obtain accurate
measurements of the host magnitude due to contaminating light from
galaxies $A$ and $B$ (Figure 9) in the ground-based images.

\begin{figure}
\begin{center} 
\includegraphics[scale=0.45,angle=0]{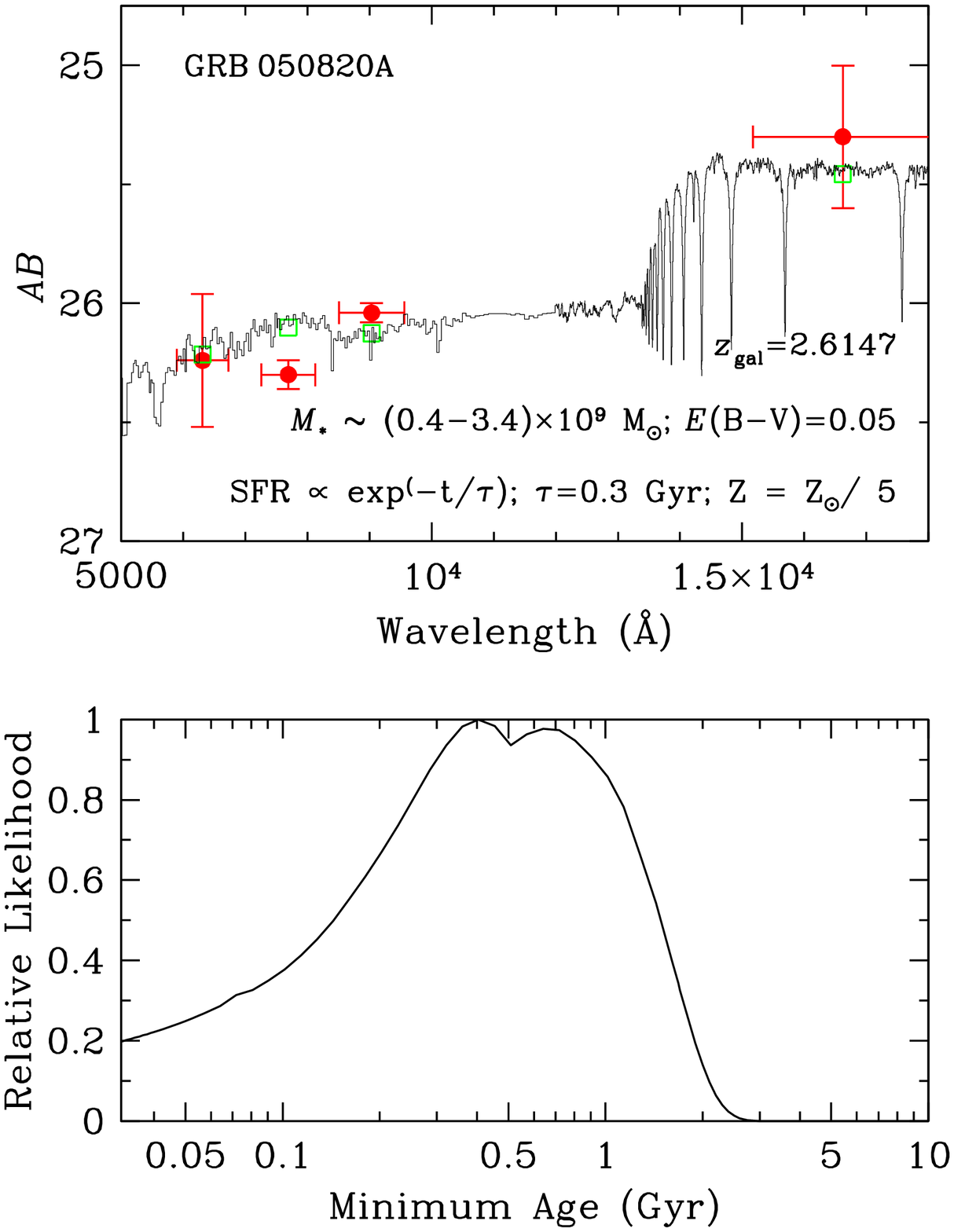}
\end{center}
\caption[]{Constraints on the galaxy age and star formation history.
Top: The broad-band spectral energy distribution of the host galaxy of
GRB\,050820A at $z=2.6147$ (points with errorbars), in comparison to
the best-fit template (solid histogram) and the predicted broad-band
photometric points (open squares).  The best-fit SED is charaterized
by a declining SFR of e-folding time $\tau=300$ Myr, $1/5$ solar
metallicity, and intrinsic dust extinction of $E(B-V)=0.05$ following
the SMC extinction law.  Bottom: The likelihood function of the
stellar age for the host galaxy, suggesting that a recent starburst
occurred about $400-700$ Myr prior to the GRB explosion.}
\end{figure}

  For the host galaxy, the optical and $H$-band photometric
measurements presented in Table 2 allow us to examine the star
formation history and constrain the stellar mass based on the observed
spectral energy distribution (SED).  To constrain the underlying
stellar population, we consider a suite of synthetic stellar
population models generated using the Bruzual \& Charlot (2003)
spectral library.  We adopt a Salpeter initial mass function with a
range of metallicity from $1/5$ solar to $2\times$ solar and a range
of star formation history from a single burst to exponentially
declining SFR of e-folding time $\tau=300$ Myr or 1 Gyr. We include
intrinsic dust extinction that follows the SMC extinction law.
Comparing the observed SED with model predictions allows us to
constrain the stellar age. The results are presented in Figure 11,
where the observed SED of the galaxy is shown in the top panel
together with the best-fit model.  It is clear that the $H$-band
photometry provides the necessary measurement for constraining the
stellar population based on the 4000-\AA\ flux decrement.  The bottom
panel of Figure 10 shows the likelihood distribution function versus
stellar age, indicating that the last major episode of star formation
occurred at $\approx 400-700$ Myr ago.

  Adopting the best-fit stellar synthetic model, we infer a total
stellar mass of $M_*= 0.9_{-0.4}^{+2.5}\times 10^9\ h^{-2}$ M$_\odot$.
We argue that our inferred stellar mass is accurate, despite an
absence of rest-frame near-infrared flux measurements.  This is
supported by previous studies, which show that stellar masses
determined based on the observed 4000-\AA\ flux decrement are
consistent with those determined based on rest-frame near-infrared
luminosity to within the uncertainties (cf.\ Chen \& Marzke 2004 and
Yan \etal\ 2004 for red galaxies identified in the Hubble Ultra Deep
Field and see also Shapley \etal\ 2005 for a detailed comparison based
on 72 star-forming galaxies at $\langle z\rangle=2.3$).  The inferred
stellar mass for the host galaxy of GRB\,050820A at $z=2.6147$ is
comparable to those derived for $z\approx 1$ GRB host galaxies (e.g.\
Castro Cer\'on \etal\ 2006, 2008; Savaglio \etal\ 2008) but falls in
the bottom 2\% of the UV luminous galaxies studied by Shapley \etal\
(2005).

\begin{center}
\begin{small}
\begin{deluxetable*}{lccccc} 
\tablewidth{0pt}
\tablecaption{A Summary of the Observed Isophotal Magnitudes of Objects at $\Delta\,\theta\le 1.5\arcsec$ from GRB\,050820A}
\tablehead{\multicolumn{1}{c}{Objects} & \colhead{$\Delta\,\theta$ (\arcsec)} & \colhead{$AB$(F625W)} & \colhead{$AB$(F775W)} & \colhead{$AB$(F850LP)} & \colhead{$AB$($H$)}}
\startdata
Host &       ...      & $26.24\pm 0.28$ & $26.30\pm 0.06$ & $26.04\pm 0.05$ & $25.3\pm 0.3$ \nl
$A$  & $1.34\pm 0.05$ & $26.08\pm 0.06$ & $26.11\pm 0.05$ & $25.81\pm 0.05$ & $>26.0$ \nl
$B$  & $0.44\pm 0.05$ & $26.35\pm 0.23$ & $26.21\pm 0.05$ & $26.25\pm 0.05$ & $>26.0$ \nl
\enddata
\end{deluxetable*}
\end{small}
\end{center}

\subsection{GRB\,050908 at $z_{\rm GRB}=3.344$}

\begin{figure}
\begin{center}
\includegraphics[scale=0.4,angle=0]{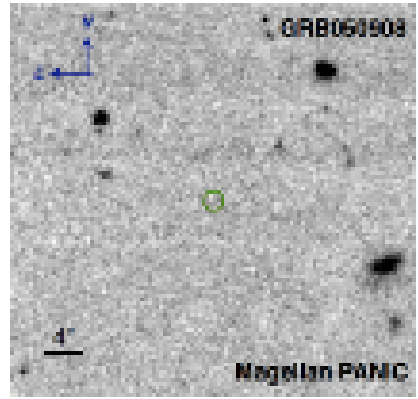}
\end{center}
\caption[]{A smoothed $H$ image of the field around GRB\,050908 at
$z_{\rm GRB}=3.344$.  No detectable flux is seen at the location of
the OT.  We place a 2-$\sigma$ $H$-band limiting magnitude of
$AB(H)=26$ over a 0.5\arcsec\ diameter aperture for the host galaxy.} 
\end{figure}

  This burst was detected by {\it Swift} (Goad \etal\ 2005) and the
optical afterglow was reported by Torii (2005) 14 minutes after the
trigger to have $R\approx 18.8$.  We obtained moderate-resolution
($\delta\,v \approx 40-150$ \kms) optical spectra of the afterglow,
using GMOS on the Gemini north telescope (Foley \etal\ 2005) and
DEIMOS on the Keck II telescope (Prochaska \etal\ 2005).  The two
spectra together provides contiguous spectral coverage over
$\lambda=5070-9000$ \AA\ (Chen \etal\ 2007b).  The afterglow spectra
exhibit a suite of absorption features, consistent with a source
redshift of $z=3.3437$.  Contrary to the majority of GRB host
galaxies, the line-of-sight toward GRB\,050908 displays only a modest
amount of neutral gas in the host galaxy.  Our initial estimate based
on the observed \lya\ absorption line suggest $\log\,N(\hI)\sim 19.2$.
However, an additional spectrum obtained by Fugazza \etal\ (2005) and
analyzed by Fynbo and collaborators displays non-negligible flux at
$\lambda_{\rm obs} < 4000$ \AA\ (corresponding to rest-frame
wavelength range $\lambda_{\rm rest} < 912$ \AA\ at $z=3.34$),
indicating $\log\,N(\hI)=17.55\pm 0.1$ (Johan Fynbo 2007, private
communication).  This source repressents one of a few GRB sightlines
observed so far that do not pass through neutral gas clouds in the
host galaxies.  Other known sources are GRB\,021004 at $z_{\rm
GRB}=2.329$ with $\log\,N(\hI)=19.5\pm 0.5$, GRB\,060526 at $z_{\rm
GRB}=3.221$ with $\log\,N(\hI)=20.00\pm 0.15$ (Jakobsson \etal\ 2006),
and GRB\,060607 at $z_{\rm GRB}=3.075$ with $\log\,N(\hI)=16.85\pm
0.10$ (see further discussion in \S\S\ 4.11 and 4.13).  A strong
Mg\,II absorber is found at $z=1.548$ in the afterglow spectrum
(Prochter \etal\ 2006).

  We have observed the field around GRB\,050908 using PANIC on
Magellan in August 2007, and obtained a total integration of 203
minutes.  The mean FWHM of the PSF was found to be $\approx
0.5\arcsec$.  The final stacked image is presented in Figure 12, which
has been smoothed using a Gaussian kernel of ${\rm FWHM}=0.5\arcsec$.
The position of the GRB is marked by a circle of 1\arcsec\ radius.  No
detectable flux is seen at the immediate location of the GRB.  We
place a 2-$\sigma$ limit in the observed $H$-band magnitude of
$AB(H)=26.0$ over a 0.5\arcsec\ diameter aperture for the host galaxy.
At $z=3.344$, the observed $H$-band magnitude limit allows us to derive a
limiting rest-frame absolute magnitude of $M_{AB}(3825) -
5\,\log\,h>-18.9$ for the GRB host galaxy.

\subsection{GRB\,050922C at $z_{\rm GRB}=2.199$}

\begin{figure}
\begin{center}
\includegraphics[scale=0.4,angle=0]{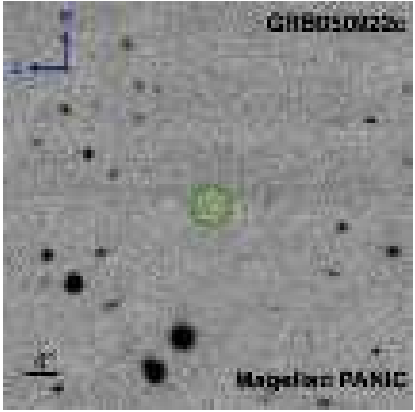}
\end{center}
\caption[]{A smoothed $H$ image of the field around GRB\,050922C at
$z_{\rm GRB}=2.199$.  No detectable flux is seen at the location of
the OT.  We place a 2-$\sigma$ $H$-band limiting magnitude of
$AB(H)=26$ over a 0.5\arcsec\ diameter aperture for the host galaxy.}
\end{figure}

  This burst was detected by {\it Swift} (Norris \etal\ 2005) and the
optical afterglow was reported by Rykoff \etal\ (2005) ten minutes
after the trigger to have $R\approx 16.0$ mag.  Spectroscopic
follow-up was carried out roughly 1.5 hours after the trigger by
Jakobsson \etal\ (2006), who reported $z_{\rm GRB}=2.198$ based on a
series of absorption features.  A foreground damped \lya\ absorber
(DLA) was also found at $z=2.07$ along this sightline.  Additional
high-resolution ($\delta\,v \approx 7$ \kms) echelle spectra of the
afterglow were obtained roughly 3.5 hours after the trigger using UVES
on the VLT Kueyen telescope (D'Elia \etal\ 2005).  The spectra were
retrieved from the ESO data archive and processed using our own
reduction software.

  The combined echelle spectrum covers a spectral range over
$\lambda=3300-10,000$ \AA, allowing accurate estimates of chemical
abudances in the ISM of the GRB host galaxy.  Based on multiple
absorption features from both ground-state and excited-state ions, we
determine a total neutral hydrogen column density of
$\log\,N(\hI)=21.5\pm 0.1$.  An absorption-line analysis of various
ions shows that $[{\rm S}/{\rm H}]= -2.03\pm 0.15$, $[{\rm Zn}/{\rm
H}]= -2.3\pm 0.3$ and $[{\rm Fe}/{\rm H}]= -2.6\pm 0.1$ (Prochaska
\etal\ 2007b; Piranomonte \etal\ 2008).  These measurements together
imply a dust-to-gas ratio roughly $1/20$ of what is seen in the SMC.
Adopting the SMC dust-to-gas ratio (Gordon \etal\ 2003), we estimate a
visual extinction in the host ISM of $A_V\approx 0.01$ (see also
Prochaska \etal\ 2007b).  Searches for H$_2$ absorption features in
the echelle spectrum has also yielded null results, placing a
4-$\sigma$ upper limit on the ISM molecular fraction of the host at
$f_{\rm H_2} < 10^{-6.8}$ (Tumlinson \etal\ 2007).  Additional strong
absorbers are found at $z=2.077$, $z=2.01$ and $z=1.99$ (Piranomonte
\etal\ 2008).

  We have observed the field around GRB\,050922C using PANIC on
Magellan in August 2007, and obtained a total integration of 209
minutes.  The mean FWHM of the PSF was found to be $\approx
0.5\arcsec$.  The final stacked image is presented in Figure 13, which
has been smoothed using a Gaussian kernel of ${\rm FWHM}=0.5\arcsec$.
The position of the GRB is marked by a circle of 2\arcsec\ radius.  No
detectable flux is seen at the immediate location of the GRB.  We
place a 2-$\sigma$ limit in the observed $H$-band magnitude of
$AB(H)=26.0$ over a 0.5\arcsec\ diameter aperture for the host galaxy.
At $z=2.199$, observed $H$-band magnitude limit allows us to derive a
limiting rest-frame absolute magnitude of $M_{AB}(B) -
5\,\log\,h>-18.2$ for the GRB host galaxy.

  Optical images of the field around GRB\,050922c have also been
obtained using LRIS and the $g$ and $R_c$ filters on the Keck I
telescope in July 2006.  The mean FWHM of the PSF was found to be
$\approx 1.2\arcsec$ in the combined $g$ image and $\approx
1.0\arcsec$ in the combined $R_c$ image.  No detectable flux is seen
at the immediate location of the GRB.  We place 2-$\sigma$ limits of
$AB(g)=27.0$ and $AB(R_c)=26.4$ over a 2\arcsec\ diameter aperture for
the host galaxy.  At $z=2.199$, observed $g$-band magnitude limit
allows us to derive a limiting rest-frame absolute magnitude of
$M_{AB}(1500) - 5\,\log\,h>-17.2$ for the GRB host galaxy.  This is
comparable to the flux limit in the $H$-band imaging data based on a
mean color of ${\rm UV}-B=1.22$ for $z=2-3$ star-forming galaxies
(Shapley \etal\ 2005).

\subsection{GRB\,060206 at $z_{\rm GRB}=4.048$}

\begin{figure}
\begin{center}
\includegraphics[scale=0.5,angle=0]{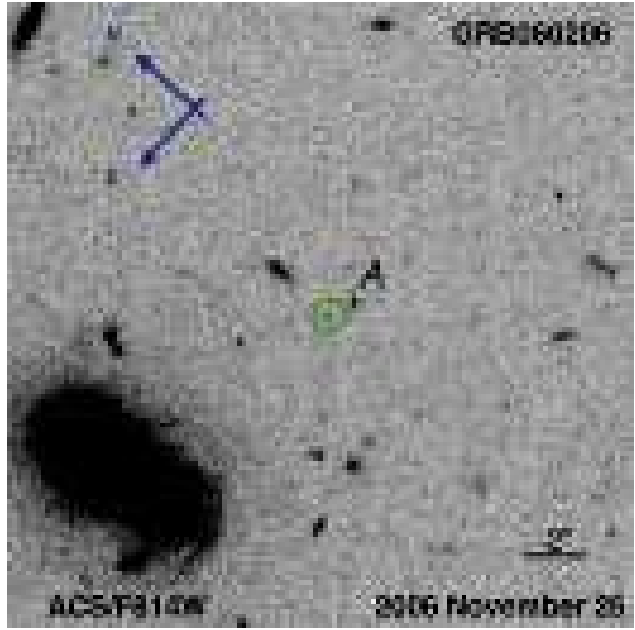}
\end{center} 
\caption[]{HST ACS/F814W image of the field around GRB\,060206 at
$z_{\rm GRB}=4.048$.  The host is detected with $AB({\rm
F814W})=27.6\pm 0.1$.  Object $A$ at $\Delta\,\theta\approx 1\arcsec$
southwest of the host has $AB({\rm F814W})=26.22\pm 0.05$ and was
initially noted by Th\"one \etal\ (2008a) as a possible host
candidate.  Given its offset from the OT position, we consider galaxy
A a foreground galaxy likely associated with one of the Mg\,II
absorbers at $z=1.48$ or $z=2.26$ found in the afterglow spectra.}
\end{figure}

  This burst was detected by {\it Swift} (Morris \etal\ 2006).  An
optical transient was nearly instantaneously identified with $V\approx
16.7$ (Fynbo \etal\ 2006b; Boyd \etal\ 2006).  Spectroscopic follow-up
of the afterglow was carried out by multiple groups (Fynbo \etal\
2006a; Prochaska \etal\ 2006; Aoki \etal\ 2006; Hao \etal\ 2007).  The
host galaxy at $z=4.048$ is found to have $\log\,N(\hI)=20.85\pm 0.10$
and $[{\rm S}/{\rm H}]= -0.85\pm 0.10$ based on moderate resolution
($\delta\,v \approx 40$ \kms) afterglow spectra (Fynbo \etal\ 2006a;
Th\"one \etal\ 2008).  No H$_2$ is detected to a 4-$\sigma$ limit of
$f_{\rm H_2} < 10^{-3.6}$ (c.f.\ Fynbo \etal\ 2006; Tumlinson \etal\
2007).  In addition to the GRB host, two strong Mg\,II absorbers are
found along the line of sight at $z=1.48$ and $z=2.26$ with
$W(2796)=0.95\pm 0.1$ \AA\ and $W(2796)=1.5\pm 0.1$ \AA, respectively
(Aoki \etal\ 2006; Hao \etal\ 2007; Th\"one \etal\ 2008).

  We obtained late-time images of the field around GRB\,060206 on
November 25, 2006, using ACS and the F814W filter on board HST
(PID$=$10817).  The images were processed and registered using the
standard pipeline technique.  A stacked ACS/F814W image is presented
in Figure 14.  At the position of the OT (marked by a circle of
0.5\arcsec\ radius), we clearly detect a faint source of $AB({\rm
F814W})=27.6\pm 0.1$.  We identify the source as the host galaxy of
GRB\,060206.  At $z=4.048$, the observed F814W magnitude corresponds
to a rest-frame absolute magnitude of $M_{AB}(1600) -
5\,\log\,h=-17.7\pm 0.1$ for the GRB host galaxy.

  At $\Delta\,\theta=0.96\pm 0.02\arcsec$ southwest of the host, we
note the presence of galaxy $A$ with $AB({\rm F814W})=26.22\pm 0.05$.
This galaxy was also seen in a deep, ground-based $r'$-band image
published in Th\"one \etal\ (2008).  The authors initially identified
galaxy $A$ as a candidate for the GRB host galaxy, but revised the
identification after analyzing the HST images.  Given the large
angular distance to the OT (corresponding to $\rho\approx 4.8\ h^{-1}$
kpc at $z=4.408$) and the presence of at least two strong Mg\,II
absorbers in the afterglow spectra, we conclude that galaxy $A$ is
likely a foreground galaxy associated with one of the Mg\,II
absorbers.

\subsection{GRB\,060607 at $z_{\rm GRB}=3.075$}

\begin{figure}  
\begin{center}
\includegraphics[scale=0.4,angle=0]{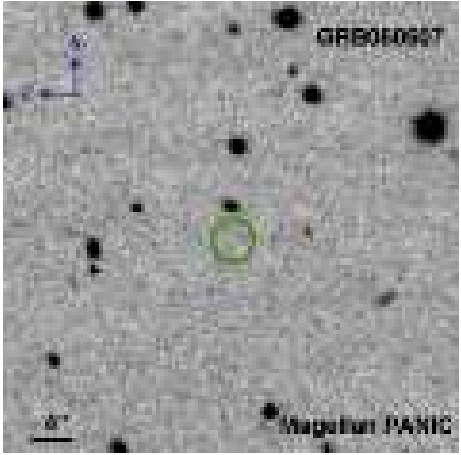}
\end{center}
\caption[]{A smoothed $H$ image of the field around GRB\,060607 at
$z_{\rm GRB}=3.075$.  No detectable flux is seen at the location of
the OT.  We place a 2-$\sigma$ $H$-band limiting magnitude of
$AB(H)=26.5$ over a 0.5\arcsec\ diameter aperture for the host galaxy.}
\end{figure}

  This burst was detected by {\it Swift} and the optical afterglow was
detected one minute after the trigger by UVOT and the white filter
($1600-6500$ \AA) on board the satellite (Ziaeepour \etal\ 2006).  The
initial brightness was estimated $\approx 15.7$ mag (Ziaeepour \etal\
2006).  A series of high-resolution ($\delta\,v \approx 7$ \kms)
optical spectra of the afterglow were obtained using UVES on the VLT
telescope by Ledoux \etal\ (2006), starting 7.5 minutes after the
trigger.  These authors identified the GRB host at $z=3.082$ and noted
two probable DLAs at $z=2.937$ and $z=3.05$.  We retrieved the echelle
spectra from the ESO data archive and processed the data using our own
reduction software.

  The afterglow spectrum of GRB\,060607 exhibits a number of unusual
features.  First, the GRB host appears to arise in a weak \lya\
absorber at $z=3.075$, with associated Si\,IV and C\,IV absorption
feature but no trace of low-ionization species.  We determine
$\log\,N(\hI)=16.85\pm 0.10$ based on a simultaneous Voigt profile
analysis of the Lyman absorption series, using the
VPFIT\footnote{http://www.ast.cam.ac.uk/\char'176rfc/vpfit.html}
software package.  This system represents the only optically thin
absorber detected in a GRB host galaxy (Jakobsson \etal\ 2006; Chen
\etal\ 2007a).  Second, the strong \lya\ absorber at $z=3.05$,
$\approx 1840$ \kms\ blueshifted from the host, is a Lyman limit
system of $\log\,N(\hI)=19.2\pm 0.1$ with weak metal absorption lines
detected, implying a metallicity $< 1/100$ solar.  Third, the strong
\lya\ absorber at $z=2.937$ exhibits complex kinematic profiles in
various ionic transitions, spreading over $\Delta\,v\approx 600$ \kms.
We measure $\log\,N(\hI)=19.6\pm 0.1$.

  We have observed the field around GRB\,060607 using PANIC on
Magellan in August 2007, and obtained a total integration of 309
minutes.  The mean FWHM of the PSF was found to be $\approx
0.5\arcsec$.  The final stacked image is presented in Figure 15, which
has been smoothed using a Gaussian kernel of ${\rm FWHM}=0.5\arcsec$.
The position of the GRB is marked by a circle of 2\arcsec\ radius.
Despite the discovery of three strong \lya\ absorbers at
$z=2.93-3.08$, no detectable flux is seen at the immediate location
of the GRB.  We place a 2-$\sigma$ limit in the observed $H$-band
magnitude of $AB(H)=26.5$ over a 0.5\arcsec\ diameter aperture for the
host galaxy.  At $z=3.075$, observed $H$-band magnitude limit allows
us to derive a limiting rest-frame absolute magnitude of $M_{AB}(B) -
5\,\log\,h>-18.3$ for the GRB host galaxy.

  Optical images of the field around GRB\,060607 have also been
obtained using LRIS and the $g$ and $R_c$ filters on the Keck I
telescope in July 2006.  The mean FWHM of the PSF was found to be
$\approx 1.0\arcsec$ in the combined $g$ image and $\approx
1.5\arcsec$ in the combined $R_c$ image.  The effective seeing was
compromised in the $R_c$ images due to a focus problem with the LRIS
red-side.  No detectable flux is found at the immediate location of
the GRB.  We place 2-$\sigma$ limits of $AB(g)=26.8$ and
$AB(R_c)=26.5$ over a 2\arcsec\ diameter aperture for the host galaxy.
At $z=3.075$, observed $R_c$-band magnitude limit allows us to derive
a limiting rest-frame absolute magnitude of $M_{AB}(1600) -
5\,\log\,h>-18.3$ for the GRB host galaxy.

\subsection{GRB\,070721B at $z_{\rm GRB}=3.626$}

\begin{figure}
\begin{center}
\includegraphics[scale=0.4,angle=0]{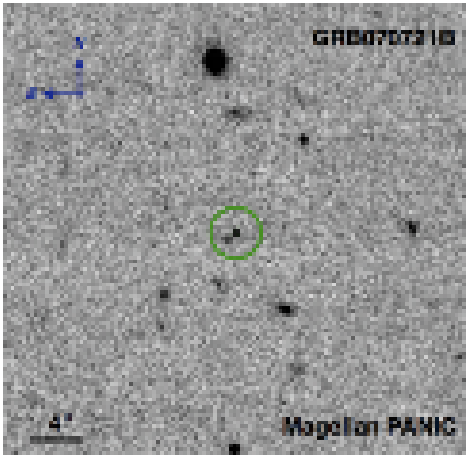}
\end{center}
\caption[]{A smoothed $H$ image of the field around GRB\,070721B at
$z_{\rm GRB}=3.626$.  No detectable flux is seen at the location of
the OT (marked by the cross).  An extended source is seen at
$\Delta\theta=0.9\arcsec$ sourtheast of the OT with $AB(H)=23.7\pm
0.1$, which has been confirmed to be at $z=3.09$ and associated with
the foreground DLA along the line of sight (Fynbo 2008, private
communication; Milvang-Jensen \etal\ 2008, in preparation).  We place
a 2-$\sigma$ $H$-band limiting magnitude of $AB(H)=25.8$ over a
0.5\arcsec\ diameter aperture for the host galaxy.}
\end{figure}

  This burst was detected by {\it Swift} (Ziaeepour \etal\ 2007) and
the optical afterglow was detected one minute after the trigger by
UVOT and the white filter ($1600-6500$ \AA) on board the satellite
(Schady \etal\ 2007).  The initial brightness was estimated $\approx
15.9$ mag (Schady \etal\ 2007).  Spectroscopic follow-up was carried
out by Malesani \etal\ (2007), who reported $z_{\rm GRB}=3.626$ based
on the presence of a strong \lya\ feature and a series of metal
absorption lines.  The host galaxy is found to have
$\log\,N(\hI)=21.50\pm 0.20$ based on moderate resolution afterglow
spectra (Jakobsson \etal\ 2008, in preparation).  A additional
intervening damped \lya\ absorber is found at $z=3.09$.

  We have observed the field around GRB\,070721B using PANIC on
Magellan in August 2007, and obtained a total integration of 172
minutes.  The mean FWHM of the PSF was found to be $\approx
0.5\arcsec$.  The final stacked image is presented in Figure 16, which
has been smoothed using a Gaussian kernel of ${\rm FWHM}=0.5\arcsec$.
The position of the GRB is marked by the cross within a circle of
2\arcsec\ radius.  No detectable flux is seen at the immediate
location of the GRB.  However, an extended source is seen at
$\Delta\theta=0.9\arcsec$ sourtheast of the OT with $AB(H)=23.7\pm
0.1$.
This source has been confirmed to be at $z=3.09$, associated with the
foreground DLA along the line of sight (Fynbo 2008, private
communication; Milvang-Jensen \etal\ 2008, in preparation).  This
represents by far the most luminous DLA galaxy found at $z>2$ (e.g.\
M{\o}ller \etal\ 2002a).  At $z=3.09$, the projected distance between
the foreground DLA and the extended source would be $\rho=4.8\ h^{-1}$
kpc.  We therefore consider the GRB host galaxy missing in our
$H$-band image and place a 2-$\sigma$ limit of $AB(H)=25.8$ over a
0.5\arcsec\ diameter aperture for the host galaxy.  At $z=3.626$,
observed $H$-band magnitude limit allows us to derive a limiting
rest-frame absolute magnitude of $M_{AB}(3600) - 5\,\log\,h>-19.3$ for
the GRB host galaxy.

\subsection{GRBs 000301C, 000926, \& 021004}

\begin{figure}
\begin{center}
\includegraphics[scale=0.32,angle=0]{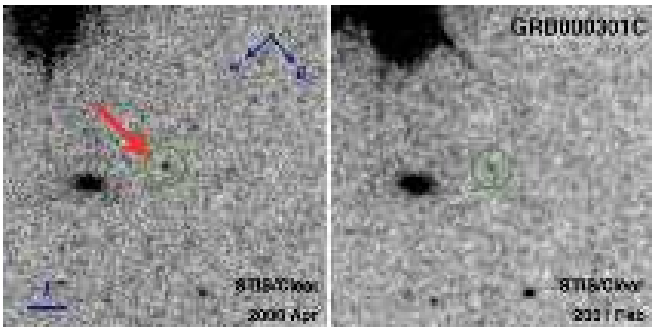}
\end{center}
\caption[]{Optical images of the field around GRB\,000301C at $z_{\rm
GRB}=2.0404$.  The epoch during which the images were taken is
indicated at the bottom of each panel.  The late-epoch image (right
panel) has been smoothed using a Gaussian kernel of ${\rm
FWHM}=0.15\arcsec$, which is roughly the size of the PSF.  At the
location of the OT, we detect faint emission of the host galaxy in the
late-time image and estimate $AB({\rm clear})=28.9\pm 0.5$ over a
0.5\arcsec\ diameter aperture (consistent with what was reported in
Fruchter \etal\ 2006).}
\end{figure}

  Early-time imaging and spectroscopic observations of the afterglow
of GRB\,000301C have been presented in Jensen \etal\ (2001), who
reported based on low-resolution afterglow spectra that the source is
at $z=2.0404$.  The host ISM in front of the afterglow was found to
have $\log\,N(\hI)=21.2\pm 0.5$.  Multi-epoch imaging follow-up was
carried out using STIS on board HST and the clear filter under program
\# 8189 (PI: Fruchter; see also Fruchter \etal\ 2006).  We retrieved
available imaging data from the HST data archive and analyzed the
images ourselves.  In the left panel of Figure 17, we present a
stacked image of the field obtained in April 2000 with a total
exposure time of 9391 s.  The optical transient is clearly visible in
this early epoch image.  In contrast, the images (total exposure time
of 7031 s) obtained in February 2001 (right panel of Figure 17)
exhibit only faint emission at the location of the GRB.  Attributing
the faint emission to the host galaxy, we measure a total flux of
$AB({\rm clear})=28.9\pm 0.5$ over a 0.5\arcsec\ diameter aperture.
This is consistent with the measurement reported by Fruchter \etal\
(2006).  Accounting for the bandpass difference (e.g.\ Figure 3), we
derive $AB(R)=28.8\pm 0.5$ and $M_{AB}(2200) - 5\,\log\,h=-15.2\pm
0.5$.

  Both GRB\,000926 and GRB\,021004 have been studied extensively by
previous authors.  Their host galaxies are identified in early effort
to carry out imaging follow-up with HST.  Here we briefly review known
emission and absorption properties of the two host galaxies to
complete the discussion of individual GRB hosts.

  Detailed spectroscopic and imaging studies of GRB\,000926 are
presented in Harrison \etal\ (2001), Fynbo \etal\ (2001), and Castro
\etal\ (2003).  The afterglow spectrum displays a DLA of
$\log\,N(\hI)=21.3\pm 0.3$ and a series of metal absorption features
at $z=2.0385$ with an estimated metallicity of $[{\rm Zn}/{\rm
H}]=-0.17\pm 0.15$ and metal abundance ratio of $[{\rm Zn}/{\rm
Fe}]=+1.3\pm 0.15$ (Castro \etal\ 2003).  Adopting the dust-to-gas
ratio and the SMC extinction law, we derive a visual extinction of
$A_V=0.15$ and $E(B-V)=0.06$.  

  The host galaxy is identified both in \lya\ emission on the ground
(Fynbo \etal\ 2002) and in space images obtained using HST WFPC2 and
the F606W filter (Castro \etal\ 2003).  It displays a disturbed
morphology at rest-frame UV wavelengths with extended emission over a
$\approx 15\ h^{-1}$ kpc projected size.  Additional near-infrared
$J$-band images show that the host has $AB(J)=24.1_{-0.4}^{+0.7}$
(Christensen \etal\ 2004) which, together with available WFPC2
photometry, constrains its absolute magnitudes at rest-frame UV and
optical wavelengths.  Including the intrinsic extinction correction
estimated from absorption line analysis, we find $M_{AB}(2000) -
5\,\log\,h=-19.63\pm 0.07$ and $M_{AB}(B) -
5\,\log\,h=-20.26_{-0.4}^{+0.7}$.  Comparing the observed SED with a
suite of stellar synthetic models described in \S\ 4.6, we find the
host galaxy is best decribed with a mean stellar age of $\approx 570$
Myr and an exponentially declined SFR of e-folding time 1 Gyr (left
panels of Figure 18).  Adopting the best-fit stellar synthetic model,
we can constrain the total underlying stellar mass to be $M_*=
2.1_{-1.9}^{+4.0}\times 10^9\ h^{-2}$ M$_\odot$, consistent with
$M_*= (1.6\pm 3.1)\times 10^9\ h^{-2}$ M$_\odot$ estimated by Savaglio
\etal\ (2008).

  Detailed spectroscopic and imaging studies of GRB\,021004 are
presented in M{\o}ller \etal\ 2002b, Mirabal \etal\ 2003, Schaefer
\etal\ 2003, Fiore \etal\ 2005, and Fynbo \etal\ 2005.  The afterglow
spectrum displays a combination of emission and absorption due to the
hydrogen \lya\ transition and a series of metal absorption lines at
$z=2.329$.  The presence of \lya\ emission in the afterglow spectrum
makes it difficult to determine $N(\hI)$ precisely.  Based on the
absence of Lyman limit photons, Fynbo \etal\ (2005) derive
$\log\,N(\hI)=19.5\pm 0.5$.  This is therefore one of the four GRB
Lyman limit absorbers published so far (including GRB\,060526 in
Jakobsson \etal\ 2006).  We have retrieved available echelle spectra
of the OT from the ESO data archive.  Based on the absorption
strengths of Si\,II and Fe\,II observed in the combined spectrum, we
take into account necessary corrections for the ionization fraction of
the gas and place an upper limit to the metallicity of the gas at
$[{\rm Si}/{\rm H}]<-1$.

  Optical and near-infrared photometry of the host galaxy of
GRB\,021004 are reported and analyzed in Fynbo \etal\ (2005).  The
authors estimate a mean stellar age of $\approx 42$ Myr.  Adopting the
optical and near-infrared photometric measurements of Fynbo \etal\
(2005), we derive a rest-frame absolute magnitude of $M_{AB}(B) -
5\,\log\,h=-20.38\pm 0.15$ or $M_{AB}(2000) - 5\,\log\,h=-19.83\pm
0.07$.  Comparing the observed SED with a suite of stellar synthetic
models described in \S\ 4.6, we find the host galaxy is best decribed
by a single starburst episode that occurred $\approx 39$ Myr ago, and
constrain the total underlying stellar mass to be $M_*=
1.3_{-0.5}^{+2.9}\times 10^9\ h^{-2}$ M$_\odot$.  This is smaller than
$M_*= (7.8\pm 0.7)\times 10^9\ h^{-2}$ M$_\odot$ estimated by Savaglio
\etal\ (2008).  The discrepancy may be understood by the poorly
constrained star formation history, displayed in the right panels of
Figure 18.

  We note that two strong Mg\,II absorbers are found at $z=1.38$ and
$z=1.60$ with $W(2796)=1.6$ \AA\ and $W(2796)=1.4$ \AA, respectively.
The large absorption width of these foreground absorbers imply
possible contaminating light in the host imaging observation (c.f.\
Pollack \etal\ 2008).  Indeed, the host galaxy is clearly resolved
into two compact clumps in available HST images (e.g.\ Fynbo \etal\
2005; Chen \etal\ 2007b), but extended \lya\ emission is detected in
ground-based narrow-band imaging follow-up (Jakobsson \etal\ 2005).
In the following discussion, we consider the derived absolute
magnitude as an upper limit to the intrinsic luminosity of this host
galaxy.

\begin{figure} 
\begin{center}
\includegraphics[scale=0.35,angle=0]{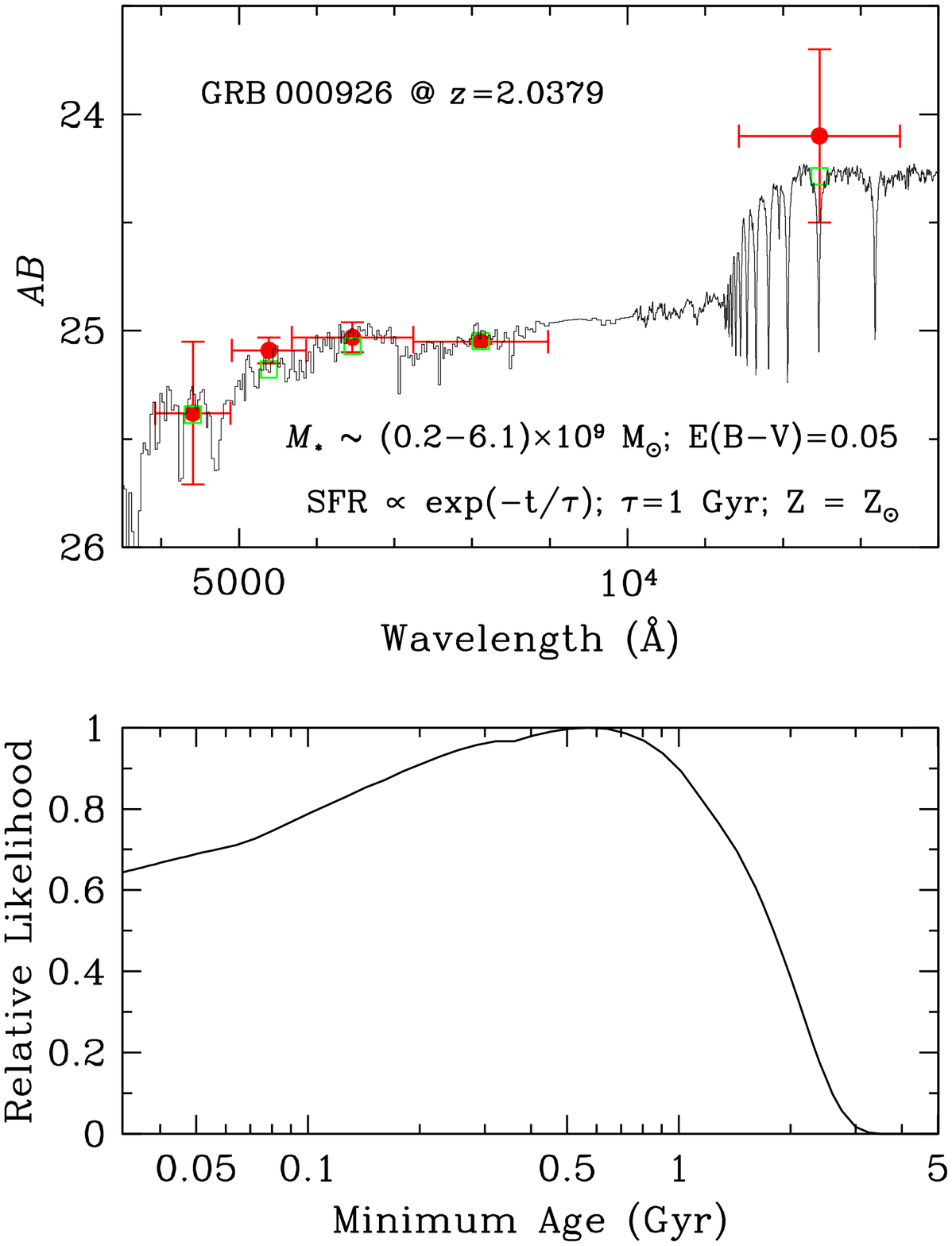}
\includegraphics[scale=0.35,angle=0]{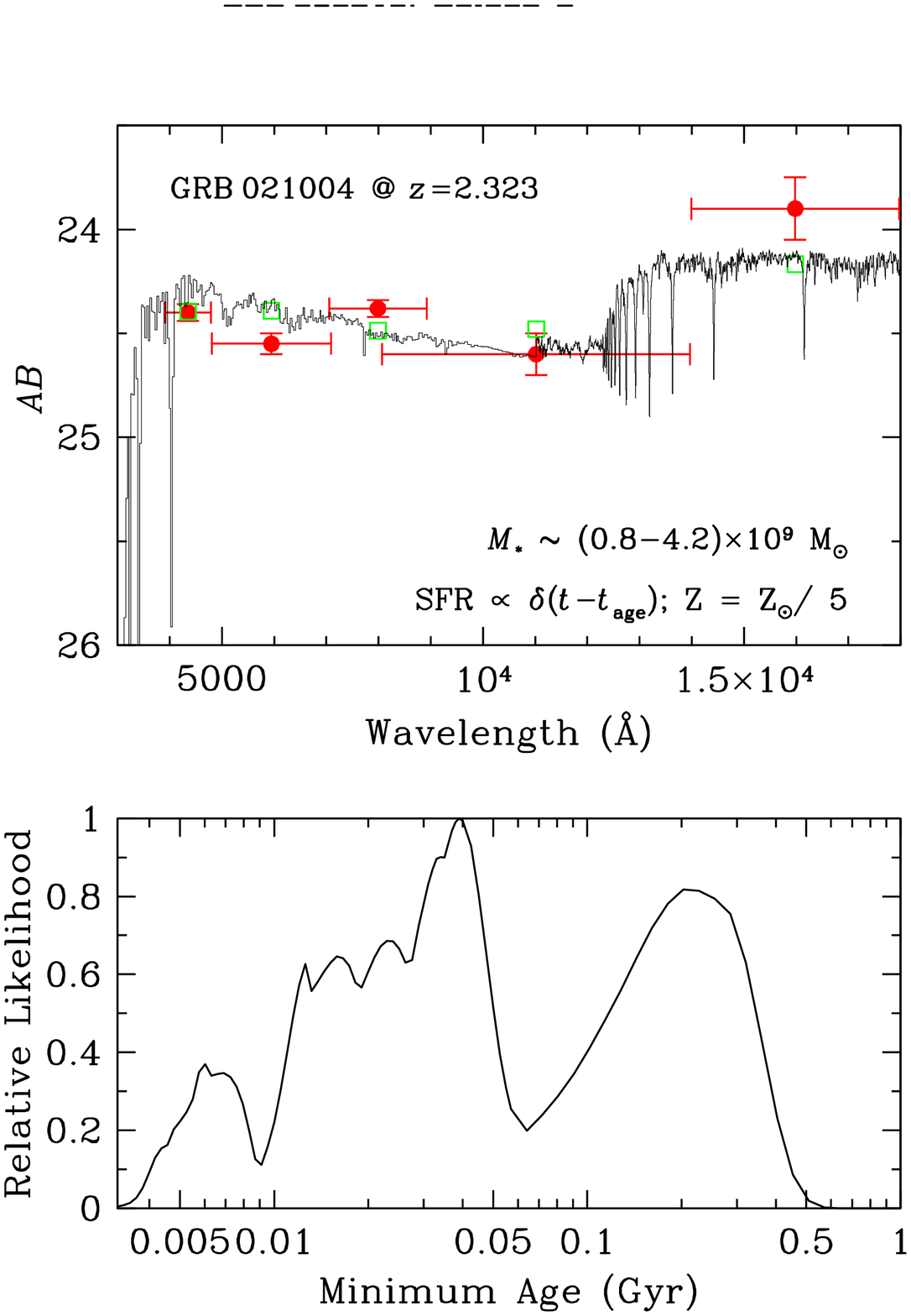}
\end{center}
\caption[]{Constraints on the galaxy age and star formation history
based on a stellar synthesis model analysis.  Left: The broad-band 
spectral energy distribution of the host galaxy of GRB\,000926 at 
$z=2.0379$ (points with errorbars), in comparison to the best-fit
template (solid histogram) and the predicted broad-band photometric
points (open squares).  The best-fit SED is charaterized by a
declining SFR of e-folding time $\tau=300$ Myr, solar metallicity, and
intrinsic dust extinction of $E(B-V)=0.025$ following the SMC
extinction law.  The likelihood function of the stellar age for the
host galaxy is display in the bottom left panel, suggesting that a
recent starburst occurred about $650$ Myr prior to the GRB explosion.
Right: A summary of the stellar synthesis model analysis for the host
galaxy of GRB\,021004 at $z=2.329$.}
\end{figure}

\section{DISCUSSION}

We have carried out an optical and near-infrared imaging survey of the
fields around 15 GRBs at $z>2$.  The GRBs are selected to have
early-time afterglow spectra in order to compare ISM absorption-line
properties with stellar properties.  The redshifts of the GRBs span a
range from $z=2.04$ to $z=4.05$, and the neutral hydrogen column
densities of the GRB host ISM span a range from $\log\,N(\hI)=16.9$ to
$\log\,N(\hI)=22.6$ (Figure 1).

In addition to the five previously known GRB host galaxies, we report
new detections for the host galaxies of GRB\,050820 and GRB\,060206.
The seven identified GRB host galaxies have rest-frame UV absolute
magnitudes spanning from $M_{AB}(U)-5\,\log\,h=-15.2$ to
$M_{AB}(UV)-5\,\log\,h=-19.8$ mag.  For sources with a known
dust-to-gas ratio from afterglow absorption-line analysis, we correct
the rest-frame UV magnitude for dust extinction assuming an SMC
extinction law.  The inferred SFR spans a range from 0.6 to 3.8
$h^{-2}$ M$_\odot$ yr$^{-1}$.  We are able to constrain the underlying
stellar populations for three host galaxies, GRB\,000926, GRB\,021004,
and GRB\,050820, based on comparisons of the observed optical and
near-infrared broad-band colors and a suite of stellar population
synthetic models .  We estimate total stellar masses of between
$M_*=(0.9-2.1)\times 10^9\ h^{-2}$ M$_\odot$ for these hosts.  We also
place 2-$\sigma$ upper limits for the rest-frame luminosities of the
remaining eight GRB host galaxies based on the depths in available
optical and near-infrared images.  Finally, high spatial resolution
images from the HST allow us to deblend the GRB host galaxies from
foreground absorbers and to unveil a range of rest-frame UV morphology
between compact (e.g.\ GRB\,060206) and extended (e.g.\ GRB\,050820)
emission features of the hosts.

A summary of known absorption-line properties and stellar properties
of the 15 GRB host galaxies is presented in Table 3.  Combining
early-time, high-resolution afterglow spectra and the results of
late-time imaging survey of the GRB fields allows us to address a
number of issues regarding both the nature of GRB progenitor
environment and star-forming physics in distant starburst galaxies.

\subsection{The Luminosity Distribution of GRB Host Galaxies}

First, we examine the luminosity distribution of GRB host galaxies
above $z=2$ based on the survey result of our sample.  The goal is to
characterize the nature of GRB host galaxies based on comparisons of
their luminosity distribution and the luminosity distribution of a
randomly selected sample from the field galaxy population.  We
incorporate the non-detections in our follow-up imaging survey by
evaluating the cumulative maximum fraction ${\cal F}_{\rm max}$ of GRB
host galaxies that are fainter than a given UV absolute magnitude
$M_{\rm max}(UV)$ (solid histogram in Figure 19).  For the host of
GRB\,030429, we have a constraint only for the rest-frame $B$-band
magnitude.  We infer its corresponding UV magnitude based on the mean
color of $\langle {\rm UV}-B\rangle=1.22$ (with an r.m.s. scatter of
0.3 mag) observed for luminous starburst galaxies at $z=2-3$ in
Shapley \etal\ (2005).  For the host of GRB\,060607, we have estimated
limiting magnitudes both in the rest-frame UV and $B$ bands.  We adopt
the more sensitive limit based on the conversion of ${\rm
UV}-B=1.22$\footnote{We note that while there is no apparent trend
between $UV-B$ and $M_{UV}$ in the luminous starburst sample of
Shapley \etal\ (2005), it is possible that that fainter dwarf
starburst galaxies may be bluer.  Given that only two fields
(GRB\,030429 and GRB\,060607) are affected by this conversion, we find
that the results presented in this section are not sensitive to the
adopted $UV-B$ color.}.  The empirical observations of the sample of
15 GRB fields confirm the previous understanding for $z\sim 1$ GRB
hosts (e.g.\ Le Floc'h \etal\ 2003): as much as 70\% of long-duration
GRBs may originate in galaxies fainter than $0.1\,L_*$.

At $z=2-4$, the field galaxy population is now well characterized by a
Schechter luminosity function $\phi(M)$ at rest-frame UV wavelengths
with $M_{AB*}-5\,\log\,h=-20.2\pm 0.1$, $\phi_*=(4\pm 0.6)\times
10^{-3}\ h^{3}\,{\rm Mpc}^{-3}$, and a faint-end slope $\alpha=-1.7\pm
0.1$ (Bouwens \etal\ 2007; Reddy \etal\ 2008).  If the GRB host
galaxies are representative of the field galaxy population, then we
expect that the fraction of host galaxies found in a luminosity
interval is proportional to the space density of galaxies in the
luminosity range.  The expected cumulative maximum fraction of the
host galaxies versus UV magnitude can be estimated following
\begin{equation}
{\cal F}_{\rm max}[M>M_{\rm max}]\propto \left [ \sum_{i=1}^{n} \phi(M_i)\,dM 
+ \sum_{j=1}^{m} \int_{M_{min}}^{M_{\rm max}} \phi(M) dM \right ],
\end{equation}
where the first term extends over $n$ known host galaxies that have
$M_{AB}(UV)>M_{\rm max}$, the second term extends over $m$
unidentified host galaxies that may be as luminous as $M_{\rm max}$,
and $\phi(M)$ is the galaxy luminosity function.  The luminosity
distribution is expected to resemble the galaxy luminosity function
with a dominant fraction attributed to faint dwarf galaxies (e.g.\
Jakobsson \etal\ 2005; Fynbo \etal\ 2008).

We test this hypothesis by calculating ${\cal F}_{\rm max}$ for a
sample of 15 random galaxies that share the known luminosities of the
seven identified GRB host galaxies and the empirical 2-$\sigma$ upper
limits for the remaining eight GRB host galaxies.  We experiment with
different $M_{\rm min}$ (with corresponding minimum luminosity $L_{\rm
min}$) in Equation (1).  The results for $L_{\rm min}=0.01\,L_*$ and
for $L_{\rm min}=0.005\,L_*$ are shown as the dashed curves in Figure
19.  Smaller $L_{\rm min}$ models predict a larger contributions from
fainter galaxies.  The model expections for different adopted $L_{\rm
min}$ clearly deviate from the observed distribution.  The hypothesis
that the host galaxies of long-duration GRBs trace random field
galaxies is rejected at $> 99$\% confidence level based on a
Kolmogorov--Smirnov (KS) test.

Next, under the assumption that GRBs trace instantaneous star
formation, galaxies with higher on-going SFR are expected to have
higher probability to host a GRB event.  Applying the rest-frame UV
luminosity as a measure of the on-going SFR, we modify Equation (1) to
include a UV luminosity weighting, $\int L\,\phi(L)\,d\,(L)$, for
calculating the expected ${\cal F}_{\rm max}$,
\begin{eqnarray}
{\cal F}_{\rm max}[M>M_{\rm max}] &\propto&
 \sum_{i=1}^{n} 10^{-0.4\,(M_i-M_*)}\,\phi(M_i)\,dM \nonumber \\ &+&
\sum_{j=1}^{m} \int_{M_{min}}^{M_{\rm max}} 10^{-0.4\,(M-M_*)}\,\phi(M) dM .
\end{eqnarray}
The results are shown as solid and dotted curves in Figure 19.  The
observations are best described under this hypothesis for $L_{\rm
min}=0.001\,L_*$ (solid curve) based on a KS test, but we cannot rule
out models with $L_{\rm min}=0.01\,L_*$ (lower dotted curve) or
$L_{\rm min}=0.0001\,L_*$ (upper dotted curve).

\begin{figure}
\begin{center}
\includegraphics[scale=0.3,angle=270]{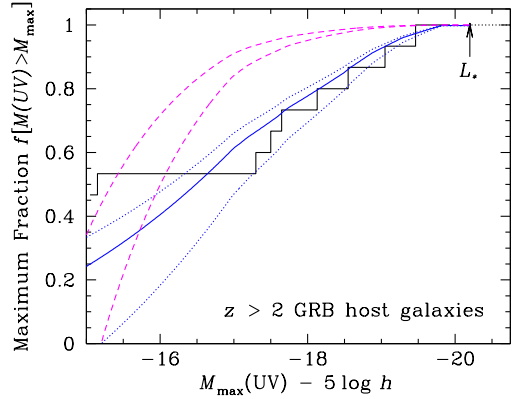}
\end{center}
\caption[]{Cumulative distribution of the maximum fraction of GRB host
galaxies that are fainter than a given UV absolute magnitude (solid
histogram), based on the seven identified host galaxies and eight
fields with upper limits in our imaging sample.  The curves show model
expectations based on different hypothesis for the origin of GRB host
galaxies.  We consider two scenarios: (1) GRB host galaxies are
representative of the general galaxy population and (2) GRB host
galaxies originate preferencially in galaxies of higher star formation
rate.  Adopting the best-fit UV luminosity function of star-forming
galaxies at $z=3-4$ from Bouwens \etal\ (2007) and Reddy \etal\
(2008), we show the expectations for scenario (1) in dashed curves for
two different minimum luminosity cutoffs, $L_{\rm min}=0.01\,L_*$
(bottom) and $L_{\rm min}=0.005\,L_*$ (top).  The expectation for
scenario (2) with $L_{\rm min}=0.001\,L_*$ is shown in the solid
curve, and the dotted curves are for $L_{\rm min}=0.01\,L_*$ (lower)
and $L_{\rm min}=0.0001\,L_*$ (upper).  The comparisons provide
empirical evidence supporting the expectation that GRB host galaxies
form an SFR weighted field galaxy sample.}
\end{figure}

In summary, the sample of seven known GRB host galaxies and eight upper
limits for unidentified hosts allows us to determine the cumulative
maximum fraction of the $z>2$ host galaxy population as a function of
rest-frame UV magnitude.  Adopting rest-frame UV luminosity as a
measure of on-going SFR, we find that the empirical sample is best
described by a SFR-weighted sample of the field galaxy population.
Models that do not include SFR weighting can be ruled out at $>99$\%
confidence level.  Based on the best-fit SFR-weighted model, we
estimate a median luminosity for the GRB host galaxies at $\approx
0.1\,L_*$.  

A similar analysis has been presented in Jakobsson \etal\ (2005), who
derived constraints for the luminosity function of GRB host galaxies
under the assumption that the observed brightness distribution of the
hosts follows a luminosity weighted field galaxy population. 
Our study differs from the approach of Jakobsson \etal\ in that our
analysis is based on a uniform set of photometric measurements from
our own imaging survey.  Then we adopt the known UV luminosity
function of the field galaxy population (with a faint-end slope of
$\alpha=-1.7$) and examine different hypotheses for generating the
observed GRB host galaxy sample.  We confirm that the host galaxy
population is representative of a UV luminosity weighted sample.  The
difference between the host luminosity function of Jakobsson \etal\
and the best-fit luminosity of star-forming galaxies at $z>2$ may be
due to uncertainties in published photometric data of the host
galaxies in their sample and uncertainties in the line-of-sight
absorption properties.

We note that although the selection criterion of our GRB sample is
based on available afterglow spectra that presumably includes only
GRBs with relatively bright optical afterglows, the broad range in the
isotropic energy release of the GRBs (see Table 1 in Chen \etal\
2007a) indicates that the GRBs in our sample are not an overly biased
portion of the long-duration GRB population.  That is, we have
selected a representative subsample of the unobscured GRB host galaxy
population.  However, the presence of 20\% dark bursts that do not
have optical afterglows found (e.g.\ Tanvir \& Jakobsson 2007)
suggests that some GRBs originate in heavily obscured star-forming
regions that are not included in our sample and are likely missed in
the UV selected field galaxy sample as well.  To constrain the
fraction of dust obscured star-forming galaxies requires an
independent study of the fields around dark bursts.

\subsection{The Luminosity--Metallicity Relation in GRB Host Galaxies}

\begin{figure}
\begin{center}
\includegraphics[scale=0.4,angle=0]{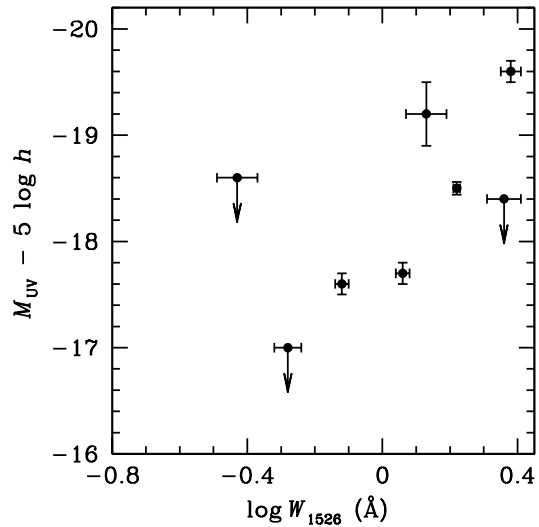}
\end{center}
\caption[]{Absolute UV magnitude of GRB host galaxies versus
rest-frame absorption equivalent width of the Si\,II $\lambda\,1526$
transition in the hosts as observed in afterglow spectra.  For this
study, we consider only host DLAs because the lyman limit absorbers do
not include neutral ISM in the host galaxies.  The Si\,II
$\lambda\,1526$ transition is typically saturated at rest-frame
absorption equivalent width $W(1526)>0.3$ \AA.  The line width
therefore provides a measure of the velocity field of cool clouds
along the line of sight through a halo (e.g.\ Prochaska \etal\ 2008).
We find a moderate trend with stronger $W(1526)$ appearing in more
luminous hosts at nearly 95\% confidence level.}
\end{figure}

We combine known absorption-line properties with estimated galaxy UV
luminosity to investigate the gas kinematics and chemical enrichment
in the ISM of GRB host galaxies.  The goals of this study are (1) to
examine the physical processes that determine the observed
absorption-properties in the host ISM, (2) to investigate whether
there is a metallicity cutoff in GRB host galaxies as favored by
various theoretical models (e.g.\ Hirschi \etal\ 2005; Yoon \& Langer
2005; Woosley \& Heger 2006), and (3) to probe the
luminosity--metallicity relation below the magnitude limit of most
previous studies at $z>2$ (e.g.\ Erb \etal\ 2006; Maiolino \etal\ 2008).

We first compare rest-frame absorption equivalent widths of the Si\,II
$\lambda\,1526$ transition $W(1526)$ observed in the host with the
absolute UV magnitude.  For this study, we consider only host DLAs
because the lyman limit absorbers do not include neutral ISM in the
host galaxies.  The Si\,II $\lambda\,1526$ transition is typically
saturated at $W(1526)>0.3$ \AA.  The line width provides a measure of
the velocity field of cool clouds along the line of sight through a
galactic halo, rather than the total gas column density (e.g.\
Prochaska \etal\ 2008).  We adopt the $W(1526)$ measurements published
in Prochaska \etal\ (2008).  Including additional measurement for
GRB\,060206 in public afterglow spectra from Subaru Science Data
Archive (Aoki \etal\ 2006), we have assembled eight GRB hosts galaxies
for this study.

Figure 20 shows that there exists a relatively significant trend with
stronger Si\,II transitions appearing in more luminous hosts.
Including upper limits, we find based on a generalized Kendall test
that the probability of a positive correlation between $M_{UV}$ and
$W(1526)$ is nearly 95\%.  While the positive correlation is similar
to the mass-metallicity relation seen in field galaxies (e.g.\ Erb
\etal\ 2006; Maiolino \etal\ 2008) , we note that half of the host galaxies have
$W(1526)\aapg 1.5$ \AA, corresponding to a velocity width of $\sim 300$
\kms.  Attributing the large line width to gravitational motion of
clouds within the host dark matter halos would require a halo mass $>
10^{11}\ h^{-1}\,{\rm M_\odot}$, comparable to the mass scale found
for halos that host luminous starburst galaxies at $z\sim 2$ (e.g.\
Conroy \etal\ 2008) but at odds with expectations for starburst
galaxies of low stellar mass and low luminosity.  But because $M_{UV}$
serves as a measure of the on-going SFR, the observed $M_{UV}$ versus
$W(1526)$ correlation implies that the velocity field observed in GRB
host galaxies is driven by galactic outflows.


Next, we examine whether there exists a correlation between host
luminosity and ISM metallicity, in comparison to what is known for
luminous starburst galaxies published by Erb \etal\ (2006).  We note
two apprarent caveats in our study.  First, chemical abundances of the
GRB host galaxies in our sample are determined for the cold neutral
medium using absorption-line techniques, whereas the metallicities of
field galaxies are determined from the integrated emission-line fluxes
of their H\,II regions.  Although extensive studies have yet to be
undertaken to examine possible systematic differences between
absorption- and emission-line abundance measurements, available
evidence based on limited studies of nearby starburst galaxies have
yielded consistent metallicity measurements using either
absorption-line or emission-line techniques (see Russell \& Dopita
1992 and Welty \etal\ 1997, 1999 for the Large and Small Magellanic
Clouds; Lecavelier des Estangs \etal\ 2004 for I\,Zw 18; and
Schulte-Ladbeck \etal\ 2005 and Bowen \etal\ 2005 for
SBS\,1543$+$593).  In the few cases where emission and absorption line
measurements have been compared in galaxies at $z > 2$, the two
methods have been found to give consistent answers to within a factor
of $\sim 2$ (e.g.\ Pettini 2006).

Second, absorption-line measurements are for the ISM along the
afterglow line of sight, whereas emission-line observations are
integrated measurements averaged over the entire galaxies.  A large
metallicity gradient is commonly seen in nearby galaxies with a slope
varying between $d\,Z/d\,r \approx -0.02$ dex / kpc to $-0.07$ dex /
kpc (e.g.\ Zaritsky \etal\ 1994; van Zee \etal\ 1998; Kennicutt \etal\
2003).  Therefore, a metallicity measurement for individual sightlines
may not be representative of the global mean value of the entire host
galaxy.  We note, however, that $z>2$ galaxies are relatively compact
with typical half light radii of $\approx 2$ kpc (Bouwens \etal\ 2004;
Law \etal\ 2007).  Previous observations (e.g.\ Bloom \etal\ 2002;
Fruchter \etal\ 2006) and current findings for GRB\,050820 and
GRB\,060206 show that the GRBs occur within $\aapl 2$ kpc radius of
their host galaxies.  This is also consistent with the large $N(\hI)$
observed in the afterglow spectra with the exception of four non-DLAs.
Spatial variation of the observed metallicity across distant
star-forming galaxies is therefore not expected to exceed 0.2 dex.  We
proceed with an analysis that compares the ISM metal content of GRB
galaxies with those of known starburst galaxies at $z\approx 2$.

Figure 21 shows the the luminosity--metallicity relation reproduced
from Erb \etal\ (2006) for starburst galaxies at $\langle
z\rangle=2.3$ (open squares).  The oxygen abundances are measured
using the $N2$ index, which is a measure of the flux ratio between the
observed [N\,II] $\lambda\,6584$ to H$\alpha$ lines\footnote{See Erb
\etal\ 2006 for a discussion of possible systematic uncertainties
associated with the $N2$ index.  While it is known that the $N2$ index
saturates at solar metallicity, the saturation does not affect our
comparison because GRB host galaxies appear to have mostly sub-solar
abundances (e.g.\ Prochaska \etal\ 2007a).}.  The rest-frame $B$-band
magnitudes published in Erb \etal\ have been converted to UV magnitude
according to $UV=B+1.22$ to facilitate comparisons with the GRB host
galaxy population and with predictions from numerical simulations
(crosses in Figure 21, reproduced from Fynbo \etal\ 2008).  The $UV-B$
color conversion is estimated based on the mean $R-K$ colors for these
galaxies in Shapley \etal\ (2005).

 We include in Figure 21 measurements for nine GRB host galaxies for
which measurments (solid points) or constraints (open circles) on the
ISM metallicity are available.  Only known host DLAs are considered in
the comparison here.  Left arrows represent the fields (GRB\,050401,
GRB\,050922c and GRB\,050730) for which no hosts have been found and
2-$\sigma$ upper limits to the intrinsic luminosity are shown.  The
lower limits represent those GRB fields for which only moderate
resolution afterglow spectra are available and the observed
absorption-line strengths represent a lower limit to the intrinsic
values (e.g.\ Prochaska 2006).

\begin{figure}
\begin{center}
\includegraphics[scale=0.4,angle=0]{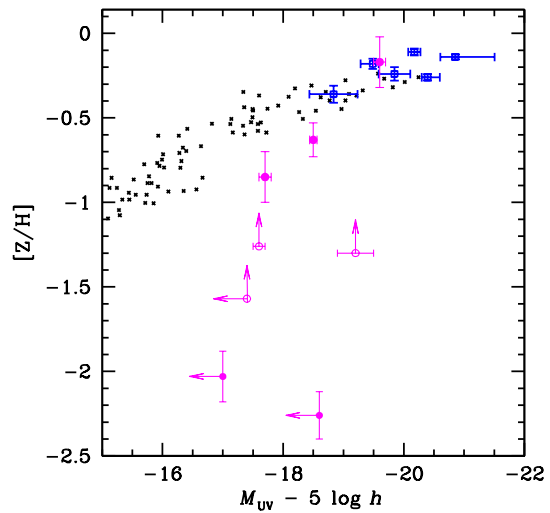}
\end{center}
\caption[]{The luminosity--metallicity relation in distant
galaxies.  Solid points with errorbars are GRB host galaxies with
accurate absorption-line metallicity measurements, while open circles
are GRB host galaxies with constraint on their ISM metallicities due to
either insufficient spectral resolution in the afterglow spectra or
unknown ionization fraction of the ISM.  Open squares are
emission-line metallicity measurements for luminous starburst galaxies
at $z\approx 2$ from Erb \etal\ (2006).  The oxygen abundance was
evaluated using the observed [N\,II] to H$\alpha$ line ratio.  We have
converted the rest-frame $B$-band magnitude to UV magnitude according
to $UV=B+1.22$, which is estimated based on the mean $R-K$ colors for
these galaxies in Shapley \etal\ (2005).  Crosses represent the
predicted luminosity--metallicity relation for starburst galaxies at
$z\sim 3$ (Fynbo \etal\ 2008) based on numerical simulations by
Sommer-Larsen \& Fynbo (2008).  }
\end{figure}

Three interesting features are seen in Figure 21.  First, the majority
of the GRB host galaxies are fainter than the faintest starburst
galaxies studied in magnitude-limited surveys (open squares) and offer
a promising probe to extend the studies of ISM metal enrichment to
fainter luminosity limits than existing faint galaxy surveys
(Djorgovski \etal\ 2004).  Second, the metallicities found in GRB host
galaxies span two orders of magnitude from $-2.2$ dex below solar to
$\approx -0.2$ below solar, while the rest-frame UV luminosity of the
galaxies extends from $\approx 1/2\, L_*$ to fainter than $0.05\,L_*$.
There is no apparent upper metallicity cutoff in the sample of GRB
host galaxies at $z>2$.  Finally, we include predictions from
numerical simulations performed by Sommer-Larsen \& Fynbo (2008) that
include SNe feedback.  The crosses are reproduced from Fynbo \etal\
(2008) for comparison with observations.  Despite the presence of
lower limits in the metallicities of three GRB hosts, there is a
moderate trend that suggests a steeper luminosity--metallicity
relation than what is seen in simulations.  We note that this is
unlikely due to a selection bias against metal enriched faint galaxies
based on the study presented in \S\ 5.1, which shows that the GRB host
galaxies are consistent with an SFR selected field galaxy sample.

Different theoretical studies have been carried out to understand the
key astrophysical processes that determine the luminosity--metallicity
relation found in galaxies from different epochs (e.g.\ Dekel \& Woo
2003; Tassis \etal\ 2008; Brooks \etal\ 2007; Finlator \& Dav\'e
2008).  The steep decline in ISM metallicity toward fainter magnitudes
may be interpreted as a signature of SNe driven outflows that removes
metals more effectively in lower-mass halos than in more massive ones
(e.g.\ Dekel \& Woo 2003), or as a signature of inefficient star
formation in low mass halos due to low gas density (e.g.\ Tassis
\etal\ 2008; Robertson \& Kravtsov 2008).  The model expectations
presented in Figure 20 includes SNe feedback but no ISM radiation
field to account for subsequent destruction of molecules in
star-forming regions.  The observed steeper luminosity--metallicity
relation in faint galaxies and the absence of molecular gas in GRB
host ISM (Tumlinson \etal\ 2007) suggests that an enhanced ISM
radiation field from young stars may be important to effectively
reduce subsequent star formation and ISM chemical enrichment in dwarf
galaxies.

\subsection{Empirical Constraints for the ISM Radiation Field in GRB Host Galaxies}

  To understand the lack of molecular gas in GRB host ISM, Tumlinson
\etal\ (2007) performed a comparison between observations and
predictions from a grid of models that cover a parameter space spanned
by four unknown properties.  These include gas density, metallicity,
clouds size and interstellar radiation field.  The authors showed that
the lack of molecular gas in the host ISM can be understood by a
combination of low metallicity (low grain production rate) and high
interstellar UV radiation field in the host ISM.  Recently, Whalen
\etal\ (2008) carried out a set of numerical simulations to examine
whether the absence of molecular gas is due to the afterglow radiation
field or an enhanced interstellar UV radiation field from a
pre-existing H\,II region where the progenitor star resides.  Taking
into account known metallicities from afterglow absorption analysis,
Whalen \etal\ concluded that, similar to what is found in the
Magellanic Clouds, ISM radiation fields of intensities up to 100 times
the Galactic mean are necessary to explain the absence of H$_2$ in the
host ISM.

  With the resolved galaxy morphologies seen in the HST images of
GRB\,000926, GRB\,030323, GRB\,050820A, and GRB\,060206, we can
measure directly the mean radiation field in the ISM of the host
galaxies at near UV wavelengths ($\approx 1500-2000$ \AA).  The mean
UV radiation intensity is determined by averaging the observed UV flux
over the extent of each host galaxy.  For comparison with models, we
convert from the observed mean radiation intensity at near UV
wavelengths to the far UV Lyman-Warner band ($11.8-13.6$ eV) based on
the Galactic UV radiation spectrum of Gondhalekar \etal\ (1980).  The
estimated far UV radiation fields $I_0$ of individual host galaxies
are presented in Table 3.  Adopting a mean Galactic radiation field at
far UV wavelengths of $I_0=2.5\times 10^{-8}$ photons cm$^{-2}$
s$^{-1}$ Hz$^{-1}$ (Gondhalekar \etal\ 1980), we find that, despite a
moderate SFR, the ISM radiation field in the GRB host galaxies spans a
range over $35 - 350 \times$ the Galactic mean value due to a
relatively compact size.  The observations provide empirical support
for the previous theoretical understanding (e.g.\ Whalen \etal\ 2008)
that strong UV radiation fields in low-metallicity ISM environment may
be a dominant factor for suppressing the formation of molecules and
therefore subsequent star formation in dwarf galaxies.

\subsection{Properties of Foreground Mg\,II Absorbing Galaxies}

  A puzzling observation of the absorption properties along GRB lines
of sight is the apparent overdensity of strong (rest-frame absorption
equivalent width $W(2796)>1$ \AA) Mg\,II absorbers (Prochter \etal\
2006).  Afterglow spectra exhibit on average $\approx 4\times$ more
strong Mg\,II absorbers than random QSO spectra (Prochter \etal\ 2006;
Sudilovsky \etal\ 2007), although such over abundance is not seen in
C\,IV absorbers at somewhat higher redshift (Sudilovsky \etal\ 2007;
Tejos \etal\ 2007).  Various scenarios have been considered to explain
this discrepancy, including dust extinction due to the presence of
these absorbers that biases observations of QSO sightlines, the GRBs
being gravitationally lensed by the foreground absorbers, and the
absorbers being intrinsic to the GRBs (Prochter \etal\ 2006; Porciani
\etal\ 2007).  However, none of these scenarios alone is found
sufficient to explain the observed overabundance along afterglow
sightlines.  A different scenario has been proposed by Frank \etal\
(2007), who consider different beam sizes between QSOs and GRB
afterglows as a possible explanation, but two important consequences
are associated with this scenario.  First, a partial covering of
Mg\,II gas is expected along QSO sightlines.  Second, a skewed
frequency distribution of Mg\,II absorbers is also expected along GRB
sightlines.  None is confirmed in empirical data (e.g.\ Pontzen \etal\
2007).  We refer the readers to Porciani \etal\ (2007) for a detailed
discussion.

  Late-time HST images of the fields around known GRBs offer a
detailed view of the galaxy environment around known foreground Mg\,II
absorbers along the sightlines toward the GRBs (e.g.\ Pollack \etal\
2008).  Six of the 15 GRBs sightlines in our sample have known strong
Mg\,II absorbers of $W(2796) > 1$ \AA\ in the foreground (e.g.\
Prochter \etal\ 2006), three of which have deep HST images available.
Including the field around GRB\,060418 (Pollack \etal\ 2008), we find
that in all four cases the HST images have uncovered at least one
galaxy at angular separation $\Delta\,\theta \aapl 1''$ from the
afterglow sightlines.  In contrast, such galaxies at small
$\Delta\,\theta$ are clearly absent in fields with no known strong
Mg\,II absorbers (Figure 22).  A summary of the known Mg\,II absorbers
and possible candidate absorbing galaxies is presented in Table 4.

\begin{figure} 
\begin{center}
\includegraphics[scale=0.35,angle=270]{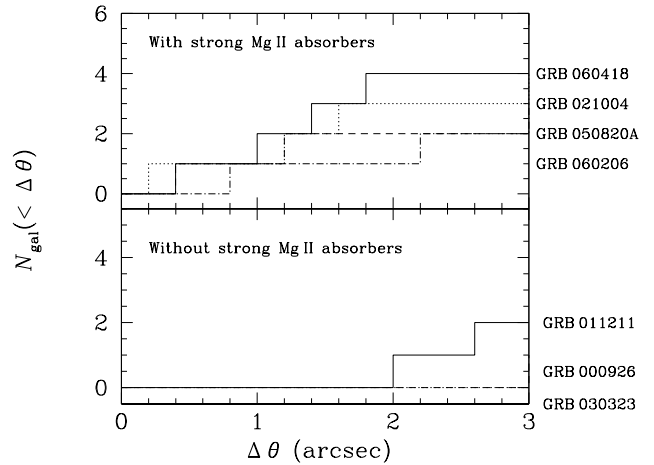}
\end{center} 
\caption[]{Cumulative number of galaxies versus angular distance to
the afterglow line of sight for GRB fields with (top) and without
(bottom) intervening strong Mg\,II absorbers found in afterglow
spectra.  All seven fields have high spatial resolution HST images
available for identifying galaxies brighter than $AB(R)=28$.  In all
four fields with known intervening strong Mg\,II galaxies, we find
additional objects at $\Delta\,\theta<2\arcsec$ from the GRB afterglow
position.  In contrast, none is seen at this small angular separation
in fields without known Mg\,II absorbers.} 
\end{figure}

  Although only one of the candidate galaxies in Table 4 has been
spectroscopically confirmed (the Mg\,II absorber at $z=0.84$ toward
GRB\,030429), the finding of additional faint galaxies along the GRB
sightlines strongly disfavors the absorbers being intrinsic to the
GRBs.  In addition, these candidate galaxies exhibit a broad range of
optical-infrared colors (c.f.\ the blue colors of galaxies toward
GRB\,050820 shown in Table 2 and Figure 9 and those found toward
GRB\,060418 in Pollack \etal\ 2008).  Dust extinction does not appear
to be a uniform factor across different GRB sightlines that results in
the higher incidence of strong Mg\,II absorbers. Comparisons of the
absorbing galaxy properties found along afterglow and QSO sightlines
(O'Meara \etal\ 2006; Becker \etal\ in preparation) should provide
further insights for understanding the differential incidences of
Mg\,II absorbers.  The morphology of additional faint emission near
the absorbing galaxy toward GRB\,030429 (Figure 6) is suggestive of a
lensed event.  Additional optical imaging and spectroscopic data are
necessary to test the lensing hypothesis for this source.

  A conclusive answer to the observed overdensity of foreground Mg\,II
absorbers along GRB sightlines requires a larger sample of imaging and
spectroscopic data of the candidate absorbing galaxies.  Based on the
current finding, however, we caution that the presence of intervening
galaxies at small angular distances to the GRBs introduces
non-negligible contamination for identifying GRB host galaxies based
on imaging data alone.  High spatial resolution images are crucial for
resolving the host galaxies from foreground absorbers.  While the
properties of the Mg\,II absorbers are beyond the scope of this paper,
we emphasize that in the absence of spectroscopic observations it
will be necessary to take into account line-of-sight absorption-line
properties in afterglow spectra when evaluating the uncertainty of an
imaging identification of GRB host galaxies.

\section{Summary}

We present a study of faint galaxies uncovered along GRB lines of
sight, based on an optical and near-infrared imaging survey of the
fields around 15 GRBs at $z>2$.  The GRBs are selected with available
early-time afterglow spectra in order to compare ISM absorption-line
properties with stellar properties.  The redshifts of the GRBs span a
range from $z=2.04$ to $z=4.05$.  The neutral hydrogen column
densities of the GRB host ISM span a range from $\log\,N(\hI)=16.9$ to
$\log\,N(\hI)=22.6$.  Our analysis differs from previous studies in
that we have obtained a uniform set of photometric data from our own
imaging survey, reducing systematic uncertainties in photometric
measurements of the host galaxies.

In addition to the five previously studied GRB host galaxies, we
consider new detections for the host galaxies of GRB\,050820 and
GRB\,060206.  We also place 2-$\sigma$ upper limits for the rest-frame
luminosities of the remaining eight GRB host galaxies based on the
depths in available optical and near-infrared images. Combining
early-time, high-resolution afterglow spectra and late-time imaging
survey of the GRB fields allows us to address a number of issues
regarding both the nature of GRB progenitor environment and
star-forming physics in distant starburst galaxies.  The line-of-sight
properties uncovered in the afterglow spectra have also proven to be
valuable for filtering potential contaminations due to foreground
galaxies.  The results of our study are summarized as the following:

  1. GRB host galaxies exhibit a broad range of rest-frame UV
absolute magnitudes spanning from $M_{AB}(U)-5\,\log\,h\aapg -15$ to
$M_{AB}(UV)-5\,\log\,h=-19.8$.  The distribution of rest-frame UV
luminosities shows that the GRB host galaxy population is best
described by a UV luminosity weighted random galaxy population with a
median luminosity of $\langle L(UV)\rangle=0.1\,L_*$.  Models that
include no luminosity weighting are ruled out at $>99$\% confidence
level.  This result demonstrates that GRB host galaxies are
representative of unobscured star-forming galaxies at $z>2$.

  2. There exists a relatively significant correlation between UV
luminosity and Si\,II $\lambda\,1526$ transition in GRB host galaxies.
A generalized Kendall test including upper limits indicates that the
probability of a positive correlation between $M_{UV}$ and $W(1526)$
is nearly 95\%.  Attributing the observed large line widths of
$W(1526)\aapg 1.5$ \AA\ to gravitational motions of gaseous clouds
would require massive halos that are also more rare at $z>2$.
Adopting $M_{UV}$ as a measure of on-going SFR and the lack of
correlation between SFR and total stellar mass in star-forming
galaxies at $z\approx 2$, we therefore interpret the observed $M_{UV}$
versus $W(1526)$ correlation as indicating that the velocity field
observed in GRB host galaxies is driven by galactic outflows.
 
  3. GRB host ISM exhibit a broad range of chemical enrichment, from
$<1/100$ solar to $\sim 1/2$ solar.  No apparent metallicty cutoff is
seen in the high-redshift host galaxy population.  Similar to nearby
galaxies, a tentative trend of declining ISM metallicity toward
fainter ($< 0.1\,L_*$) luminosities is seen in the star-forming galaxy
population at $z=2-4$.  The slope is steeper than what is expected in
some numerical simulations that incorporate supernovae feedback.
Together with an absence of molecular gas and the presence of large
amounts atomic gas, the observed luminosity--metallcity relation may
be explained by low star formation efficiency in dwarf galaxies.

  4. We measure the interstellar radiation field using resolved
rest-frame UV morphologies in available HST images.  We find that the
UV radiation field in GRB host ISM spans a range over $\approx
35-350\times$ higher than the Galactic mean value.  The strong ISM
radiation field observed in GRB host galaxies is expected to increase
the formation threshold of molecules and suppress subsequent star
formation, supporting the hypothesis that star formation efficiency in
dwarf galaxies is reduced due to the strong radiation field of
existing H\,II regions.

  5. We examine the galaxy environment in available HST images of
strong Mg\,II absorbers found along GRB sightlines.  In all fields
with known strong absorbers, we identify at least one faint galaxy at
$\aapl 1''$ from the afterglow position.  In contrast, such galaxies at
small $\Delta\,\theta$ are clearly absent in fields with no known
strong Mg\,II absorbers.  The finding of additional faint galaxies
along the GRB sightlines strongly disfavors the strong absorbers being
intrinsic to the GRBs.  Additional imaging and spectroscopic data of
the candidate absorbing galaxies are necessary to investigate the
effect of dust and gravitational magnification, but the presence of
intervening galaxies at small angular distances to the GRBs increases
the ambiguity of identifying GRB host galaxies.  It is necessary to
combine high spatial resolution images and early-time afterglow
spectra for accurate identifications of the host galaxies.

\acknowledgments

  The authors thank N. Gnedin, R. Kennicutt, B. M\'enard,
E. Ramirez-Ruiz, C.  Thom for important discussions.  We also thank
J.\ Fynbo and P.\ Jakobsson for constructive comments that helped
improve the presentation of the paper.  Support for the HST program
\#10817 was provided by NASA through a grant from the Space Telescope
Science Institute, which is operated by the Association of
Universities for Research in Astronomy, Inc., under NASA contract NAS
5-26555.  H.-W.C. acknowledges partial support from an NSF grant
AST-0607510 and HST-GO-10817.01A.

\newpage

\begin{landscape}
\begin{deluxetable}{lcccccrcccccc}
\tabletypesize{\small}
\tablewidth{0pt}
\tablecaption{Summary of known ISM and Stellar Properties of GRB Host Galaxies at $z>2$} 
\tablehead{ \colhead{} & \colhead{} & \multicolumn{6}{c}{Absorption Properties} & \colhead{} & \multicolumn{4}{c}{Galaxy Properties} \\ 
\cline{3-8} 
\cline{10-13} \\ 
\colhead{} & \colhead{} &\colhead{} & \colhead{} & \colhead{$W(1526)^a$} & \colhead{} & \colhead{} & \colhead{} & \colhead{} & \colhead{$M_{AB}(UV)$} & \colhead{$I_0^e$} & \colhead{$M_{AB}(B)^f$} & \colhead{$M_*$} \\
\colhead{Field} & \colhead{} &\colhead{$z_{\rm GRB}$} & \colhead{$\log\,N(\hI)$} & \colhead{(\AA)} & \colhead{$\left [\frac{\rm M}{\rm H}\right ]_{\rm ISM}^b$} & \colhead{$\left [\frac{\rm M}{\rm Fe}\right ]_{\rm ISM}^c$} & \colhead{$A_V^d$} & \colhead{} & \colhead{$-5\,\log\,h$} & (photons/cm$^2$/s/Hz) & \colhead{$-5\,\log\,h$} & \colhead{($\times 10^9\ h^{-2}$ M$_\odot$)}}
\startdata
GRB\,000301C\dotfill  & & 2.040 & $21.2 \pm 0.5$   &       ...       &       ...        &       ...        &   ... & & $-15.2\pm 0.5$ & ... & ... & ... \nl
GRB\,000926 \dotfill  & & 2.038 & $21.3 \pm 0.3$   & $2.40\pm 0.14$  & $-0.17\pm 0.15$ &  $+1.32\pm 0.15$ &  0.15 & & $-19.63\pm 0.07$ & $3.3\times 10^{-6}$ & $-20.26_{-0.4}^{+0.7}$ & $2.1_{-1.9}^{+4.0}$ \nl
GRB\,011211 \dotfill  & & 2.140 & $20.4 \pm 0.2$   & $1.35\pm 0.19$  &  $> -1.3$        &       ...        &   ... & & $-19.2\pm 0.1$ & ... & $-19.1\pm 0.3$ & ... \nl
GRB\,021004 \dotfill  & & 2.329 & $19.5 \pm 0.5$   &      ...        & $< -1.0$        &       ...        &   ... & & $-19.83\pm 0.07$ & ... & $-20.38\pm 0.15$ & $2.1_{-1.9}^{+4.0}$ \nl
GRB\,020124 \dotfill  & & 3.198 & $21.7 \pm 0.2$   &      ...        &      ...        &       ...        &   ... & & $>-15.1$ & ... & ... & ... \nl
GRB\,030323 \dotfill  & & 3.372 & $21.90 \pm 0.07$ & $0.75\pm 0.03$  & $> -1.26$       &       ...        &   ... & & $-17.6\pm 0.1$ & $8.8\times 10^{-7}$ & 18.8 & ... \nl
GRB\,030429 \dotfill  & & 2.658 & $21.6 \pm 0.2$   &       ...        &     ...        &       ...        &   ... & & ... & ... & $>-20.1$ & \nl
GRB\,050401 \dotfill  & & 2.899 & $22.6 \pm 0.3$   & $2.31\pm 0.26$  & $> -1.57$       &       ...        &   ... & & $>-17.4$ & ... & $>-19.6$ & ... \nl
GRB\,050730 \dotfill  & & 3.968 & $22.15 \pm 0.10$ & $0.37\pm 0.05$  & $-2.26\pm 0.14$ &  $+0.24\pm 0.11$ &  0.00 & & $>-18.6$ & ... & ... & ... \nl
GRB\,050820A \dotfill & & 2.615 & $21.0 \pm 0.1 $  & $1.65\pm 0.05$  & $-0.63\pm 0.11$ &  $+0.97\pm 0.15$ &  0.08 & & $-18.50\pm 0.06$ & $5.6\times 10^{-6}$ & $-19.3\pm 0.3$ & $0.9_{-0.4}^{+2.5}$ \nl
GRB\,050908 \dotfill  & & 3.343 & $17.55 \pm 0.10$ &      ...        &      ...        &       ...        &   ... & & $>-18.9$ & ... & ... & ... \nl
GRB\,050922C \dotfill & & 2.199 & $21.5 \pm 0.1$   & $0.52\pm 0.05$  & $-2.03\pm 0.15$ &  $+0.60\pm 0.10$ &  0.01 & & $>-17.1$ & ... & $>-18.2$ & ... \nl
GRB\,060206 \dotfill  & & 4.048 & $20.85 \pm 0.10$ & $1.15\pm 0.05$  & $-0.85\pm 0.15$ &       ...        &   ... & & $-17.7\pm 0.1$ & $9.1\times 10^{-6}$ & ... & ... \nl
GRB\,060607 \dotfill  & & 3.075 & $16.85 \pm 0.10$ &      ...        &      ...        &       ...        &   ... & & $>-18.3$ & ... & $>-18.3$ & ... \nl
GRB\,070721B \dotfill & & 3.626 & $21.50\pm 0.20$  &       ...        &     ...        &       ...        &   ... & & $>-19.3$ & ... & ... & ... \nl
\enddata 

\tablenotetext{a}{Rest-frame absorption equivalent width of Si\,II
$\lambda1526$ in the host ISM.  The measurements of all but the one
for GRB\,060206 are adopted from Prochaska \etal\ (2008).  The
measurement for GRB\,060206 is determined using public spectra in
Subaru Science Data Archive (Aoki \etal\ 2006). }
\tablenotetext{b}{All measurements but the one for GRB\,000926 are
based on the observed S\,II absorption strength.  The measurement for
GRB\,000926 is based on the observed Si\,II absorption strength,
because S\,II transitions are not covered in available moderate
resolution spectra of the afterglow.  All lower limits are based on
the observed Zn abundance in low-to-moderate resolution afterglow
spectra. The upper limit for GRB\,021004 marks the maximum Si
abundance prior to an ionization fraction correction.}
\tablenotetext{c}{The relative abundances are measured based on sulfur
wherever sulfur is available.  For GRB\,00096, the reported value is
based on silicon. }  \tablenotetext{d}{The values are derived,
assuming the SMC extinction law (Gordon \etal\ 2003).}
\tablenotetext{e}{ISM far UV radiation field estimated from resolved
host galaxy images in HST data.}  \tablenotetext{f}{The rest-frame
$B$-band absolute magnitude is derived based on available $H$-band
photometry in the observed frame.  For sources at $z>3$ with detected
emission, we infer $M_{AB}(B)$ from the measured $M_{AB}(UV)$ and the
mean color of starburst galaxies at $z=2-3$ from Shapeley \etal\
(2005).}
\end{deluxetable}
\clearpage
\end{landscape}

\begin{deluxetable*}{p{1in}ccccrccc}
\tabletypesize{\scriptsize} 
\tablewidth{0pt}
\tablecaption{Summary of Candidate Galaxies Associated with Foreground $W(2796)>1$ \AA\ Mg\,II Absorbers} 
\tablehead{ \colhead{} & \colhead{} & \colhead{} & \multicolumn{2}{c}{Mg\,II Properties} & \colhead{} & \multicolumn{3}{c}{Galaxy Properties} \\ 
\cline{4-5} 
\cline{7-9} \\ 
\colhead{Field} & \colhead{} &\colhead{$z_{\rm GRB}$} & \colhead{$z_{\rm Mg\,II}$} & \colhead{$W(2796)$\tablenotemark{a}} & \colhead{} & \colhead{$AB$} & \colhead{$\Delta\,\theta (\arcsec)$} & \colhead{Filter}}
\startdata
GRB\,021004 \dotfill  & & 2.329 & 1.38 & 1.81 & & $> 24.4$ & $\apl 0.3$ & F606W \nl
                      & &       & 1.60 & 1.53 & & $> 24.4$ & $\apl 0.3$ & F606W \nl
GRB\,030429 \dotfill  & & 2.658 & 0.84 & 3.30 & & $20.57\pm 0.05$ & $\approx 1.3''$ & $H$ \nl
GRB\,050820A \dotfill & & 2.615 & 0.69 & 2.99 & & $26.30\pm 0.05$ & $\approx 0.4''$ & F625W \nl
                      & &       & 1.43 & 1.89 & & $26.20\pm 0.05$ & $\approx 1.3''$ & F625W \nl
GRB\,060206 \dotfill  & & 3.548 & 2.26 & 1.60 & & $26.22\pm 0.55$ & $\approx 1.0''$ & F814W \nl
GRB\,070721B \dotfill & & 3.626 & 3.09  & ... & & $23.7\pm 0.1$ & $\approx 0.9''$ & $H$ \nl
\enddata 

\tablenotetext{a}{Measurements of Mg\,II absorbers along GRB\,021004,
GRB\,030429, and GRB\,050820A are adopted from Prochter \etal\ (2006).
The Mg\,II absorber toward GRB\,060206 is identified based on our own
analysis of available FOCAS spectra.  The strong absorber toward
GRB\,070721B is identified based on a DLA feature by Fynbo and
collaborators (private communication).}

\end{deluxetable*}

\end{document}